\documentclass[11pt,twoside,a4paper,dvipsnames]{article}

\usepackage{chemformula} % Formula subscripts using \ch{}
\usepackage[T1]{fontenc} % Use modern font encodings
\usepackage{graphicx}% Include figure files
\usepackage{dcolumn}% Align table columns on decimal point
\usepackage{bm}% bold math
\usepackage[utf8]{inputenc}
\usepackage{amsthm,amscd}
\usepackage{braket}
\usepackage{paralist}
\usepackage{booktabs}
\usepackage{color}
\usepackage{mathrsfs}
\usepackage{subcaption}
\usepackage{siunitx}
\usepackage{xfrac}
\usepackage{xcolor}
\usepackage{tikz}

\usepackage{geometry}
\usepackage{tikz}\usepackage[sorting=none  , 
style=phys,
maxbibnames=99,
autocite=superscript]{biblatex} 
\bibliography{Lit}% Produces the bibliography via BibTeX.

\newcommand{\tn}{\textnormal}
\newcommand{\BB}{\ensuremath{\ \tn{B}_0}}
\renewcommand{\eqref}[1]{Eq.~(\ref{#1})}

\newcommand{\mEh}{\ensuremath{\ \tn{m}E_\tn{h}}}
\renewcommand{\AA}{\ensuremath{\ \tn{\r{A}}}}
\newcommand{\EhB}{\ensuremath{\ E_{\tn{h}}/\tn{B}_{0}}}
\newcommand{\figref}[1]{Fig.~\ref{#1}}

\newcommand{\Bohr}{\ensuremath{\ a_0}}

\newcommand{\onlinecite}[1]{\cite{#1}}

\begin{document}
\author{Marios-Petros Kitsaras$^{1,2,3}$, Florian Hampe$^{2,4}$, Lena Reimund$^{2}$, Stella Stopkowicz$^{3,5,4,2}$}
\date{$1$ Laboratoire de Chimie et Physique Quantiques - UMR5626, CNRS, Université de Toulouse - Bat. 3R1b4 - 118 route de Narbonne, F-31062, Toulouse, France \\
      $2$ Department Chemie, Johannes Gutenberg-Universit\"at Mainz, Duesbergweg 10-14, D-55128 Mainz, Germany \\
      $3$ Fachrichtung Chemie, Universit{\"a}t des Saarlandes, D-66123 Saarbr{\"u}cken, Germany \\
      $4$ Centre for Advanced Study (CAS) at the Norwegian Academy of Science and Letters, Drammensveien 78, N-0271 Oslo, Norway \\
      $5$ Hylleraas Centre for Quantum Molecular Sciences, Department of Chemistry, University of Oslo, P.O. Box 1033 Blindern, N-0315 Oslo, Norway}

\title{Equation-of-motion coupled-cluster variants in combination with perturbative triples corrections in strong magnetic fields}

\maketitle

\begin{abstract}

In this paper, we report on the implementation of the EOM spin-flip (SF), ionization-potential (IP) and electron-affinity (EA) coupled cluster singles doubles (CCSD) methods for atoms and molecules in strong magnetic fields for energies as well as one-electron properties. 
Moreover, non-perturbative triples corrections using the EOM-CCSD(T)(a)* scheme were implemented in the finite-field framework for the EE, SF, IP, and EA variants. 
These developments allow the access to a large variety of electronic states as well as the investigation of IPs and EAs in a strong magnetic field. The latter two indicate the relative stability of the different oxidation states of elements. The increased flexibility to target challenging electronic states and the access to the electronic states of the anion and cation are important for the assignment of spectra from strongly magnetic White Dwarf (WD) stars. Here, we investigate the development of the IPs and EAs in the presence of  a magnetic field for the elements of the first and second row of the periodic table. In addition, we study the development of the electronic structure of Na, Mg, and Ca that aided in the assignment of metal lines in a magnetic WD. Lastly, we investigate the electronic excitations of CH in different magnetic-field orientations and strengths, a molecule that has been found in the atmospheres of WD stars.

\end{abstract}

\section{Introduction}
%Best introduction ever here.
Chemical investigations of astronomical interest form the field of astrochemistry.\cite{Millar2015,Jorgensen2020} While this branch is typically concerned with the interstellar medium, studying the chemical composition of celestial objects like stars is key in deciphering the stellar evolution.\cite{Straniero2003} Among these investigations, the study of highly magnetic White Dwarfs (WDs), is particularly challenging.\cite{Weidemann1968,Jordan2008,Deglinnocenti2004,Ferrario2015,Bagnulo2020} These stellar remnants may exhibit magnetic fields at the order of $\sim 1\BB$. Magnetic fields of this magnitude are not reproducible on Earth and, as such, these condition are difficult to  model in an experimental setting.\cite{Murdin2013} In the absence of reference data, high-quality theoretical predictions are needed for the interpretation of spectra. Such predictions are however still rather limited due to the fact that a non-perturbative treatment is required which is still non-standard. The magnetic interactions compete with the electrostatic ones for field strengths of this magnitude\cite{Schmelcher1997,Becken1999,Al-Hujaj2004,Schmelcher2012} giving rise to complex electronic structures and chemical phenomena without a field-free analogue. Examples include the perpendicular paramagnetic bonding mechanism\cite{Lange2012} as well as exotic molecular structures.\cite{Pemberton2022} The fact that  molecules have been observed on non-magnetic or slightly magnetic WDs\cite{Berdyugina2007,Xu2013} further drives theoretical investigations to the study of molecules in this so-called "mixing" regime. 

As the magnetic interactions are at the same order of magnitude as the electrostatic ones, the use of finite magnetic-field methods (ff) rather than a perturbative treatment is required. Typical challenges in this ff framework are: 1) dealing with the gauge-origin problem, 2) the need for complex algebra, and 3) the resulting increase in computational cost. The first challenge has been addressed by employing London orbitals\cite{London1937} that ensure gauge-origin independent energies and observables for approximate wavefunctions. The first implementation of a complex ff Hartree-Fock Self-Consistent-Field (HF-SCF) method for the study of molecules in an arbitrary orientation of the magnetic field has been presented in Ref.~\onlinecite{Tellgren2008}. Since then, various ff developments of quantum-chemical methods, as for example, self-consistent field methods,\cite{Sun2019,Sun2019b,Lehtola2020,Bischoff2020,David2021} current density-functional theory,\cite{Tellgren2014,Pausch2022,Irons2021,Pemberton2022} semi-empirical methods,\cite{Cheng2023} coupled-cluster methods,\cite{Stopkowicz2015,Hampe2017,Hampe2019,Hampe2020,Blaschke2021,Culpitt2023_un,Kitsaras2024,Blaschke2024} Green's-function approaches,\cite{Holzer2019,Holzer2021}, explicit consideration of non-uniform magnetic fields,\cite{Sen2018,Sen2019} molecular dynamics,\cite{Culpitt2021,Peters2021,Monzel2022} time-dependent approaches,\cite{Ofstad2023,Culpitt2023_thf} methods for the treatment of molecular vibrations\cite{Tellgren2023} and non-adiabatic couplings\cite{Culpitt2024} have been presented. As for the third challenge,  efficient implementations and cost reducing strategies are needed. Apart from the use of density fitting\cite{Reynolds2015,Pausch2020} Recently,  approximations to the standard Coupled-Cluster (CC) truncations\cite{Kitsaras2024} and the use of Cholesky decomposition\cite{Blaschke2021, Blaschke2024} in ff investigations have been employed. 
Today, there are several quantum-chemical programs available in which various ff methods are implemented.\cite{qcumbre,cfour,LONDON,quest,bagel,WilliamsYoung2020,TURBOMOLE,Bischoff2020,Lehtola2020}

Among  different approaches in quantum-chemistry, the CC approach\cite{Cizek1966,Monkhorst1977,Shavitt2009} has proven instrumental in the study of highly magnetic WDs\cite{Stopkowicz2015,Hampe2017,Stopkowicz2018} and has recently lead to the first assignment of metals in a strongly magnetic WD.\cite{Hollands2023} The high accuracy that this approach can achieve is key for theoretical predictions that can be used for the assignment of spectra.\cite{Hampe2020} For such assignments, transition wavelengths as well as intensities as functions of the magnetic field strength are required. Within the CC framework, these are accessible via the equation-of-motion (EOM-)CC approach and the prediction of field-dependent excitation energies\cite{Hampe2017} and transition moments.\cite{Hampe2019} Beyond the standard excitation-energy (EE)-EOM-CC formulation, the EOM ansatz can also be used for the targeting of states with different multiplicities via the spin-flip (SF) variant,\cite{Levchenko2004} or states with a different number of electrons via the ionization-potential (IP)\cite{Stanton1994} and electron-affinity (EA) variants.\cite{Nooijen1995} For example, such approaches enable the access to the triplet manifold starting from a singlet-state reference, the investigation of IPs and EAs, respectively. In addition, these EOM-CC flavours also facilitate the targeting of states with a challenging electronic structure. For example, SF-EOM-CC can be employed in the study of biradicals,\cite{Krylov2006} where the additional inclusion of approximate triples excitation leads to high-quality results.\cite{Verplancke2023}  Additionally,  open-shell states with possible multiconfigurational character can be targeted using a well-behaved single-determinant reference state.\cite{Krylov2008,Krylov2017} This flexibility is invaluable when studying exotic electronic structures in the presence of a magnetic field, as their character may change drastically in different magnetic-field strengths and orientations.\cite{Hampe2020,Kitsaras2024}

The assignment of electronic spectra typically requires an accuracy beyond that of CC singles doubles (CCSD). Furthermore, in the presence of a magnetic field, the varying character of states in different strengths and orientations may drastically  influence the accuracy of the prediction when a predominant double-excitation character is present within the state of interest.\cite{Hampe2020} These shortcomings may be remedied at the CC singles doubles triples (CCSDT) level of theory, but the $M^8$  scaling of the method, where $M$ is the number of basis functions, significantly limits the applicability of the approach. The gold standard CCSD(T)\cite{Raghavachari1989} with its non-iterative $M^7$ corrections only partially addresses this issue, as it is not applicable for excited states at the EOM-CC level. The CC3 approximation\cite{Christiansen1995,Koch1997} has merit for ff investigations,\cite{Kitsaras2024} but its iterative $M^7$ scaling may also prove non feasible for larger systems. While a standard for non-iterative triples corrections at the EOM-CC level is not established in the literature, the EOM-CCSD(T)(a)* approach\cite{Matthews2016} has been shown to give balanced results.\cite{Matthews2020a}

In this paper, we report on the implementation of ff variants of the  SF-, IP-, and EA-EOM-CCSD methods for energies and one-electron properties at the expectation-value level of theory.\cite{Stanton1993,Hampe2019} In addition, we present an implementation of the ff-CCSD(T)(a)* approach. The approach consists of the CCSD(T)(a) perturbative correction for the reference state, and the so-called star (*) correction for the EOM-CC state. The use of this approach in combination with the EA, IP and SF variants is reported in this work for the first time for both the field-free and the finite fied case. The methods are used to study the IPs and EAs of the lighter main group elements in the presence of a magnetic field. 

In addition, the evolution of the electronic structure of metal atoms in strong magnetic fields is investigated: The IPs of Na and Mg as well as the electronic excitations of Ca are simulated using the implemented ff-EOM-CCSD(T)(a)* variants which were partly used for the assignment of spectra from a magnetic WD star.\cite{Hollands2023}  Lastly, the increased flexibility of the implemented EOM-CC flavors is tested for low-lying excited states of the CH radical, a molecule occurring on WD stars, which is a challenging case for the ff EE-EOM-CC approach.\cite{Kitsaras2024}

%%%%%%%%%%%%%%%%%%%%%%%%%%%%%%%%%%%%%%%%%%%%%%%%%%%%%%%%%%%%%%%%%%%%%%%%%%%%
\section{\label{Theory}Theory}
%%%%%%%%%%%%%%%%%%%%%%%%%%%%%%%%%%%%%%%%%%%%%%%%%%%%%%%%%%%%%%%%%%%%%%%%%%%%
\subsection{Molecular Hamiltonian in the presence of a uniform magnetic field}
In a uniform magnetic field, the electronic Hamiltonian for an $N$-electron molecule is
\begin{align}
\hat{H}&=\hat{H}_0+\frac{1}{2}\sum\limits_i^N\bm{B}\cdot\hat{\bm{l}}_i^{\bm{O}}+\sum\limits_i^N\bm{B}\cdot\hat{\bm{s}}_i+\frac{1}{8}\sum\limits_i^N\left(B^2r_i^{\bm{O}^2}-\left(\bm{B}\cdot\bm{r}_i^{\bm{O}}\right)^2\right).
\label{Hamiltonian}
\end{align}
$\hat{H}_0$ is the field-free molecular Hamiltonian. $\bm{B}$ denotes the vector of the magnetic field and $\hat{\bm{s}}_i$ the spin of electron $i$, $\bm{r}_i^{\bm{O}}$ its position vector with respect to the gauge origin $\bm{O}$, and $\hat{\bm{l}}_i^{\bm{O}}=-\text{i}\bm{r}_i^{\bm{O}}\times\bm{\nabla}_i$ the canonical angular momentum operator. The contribution that scale linearly with the magnetic field are referred to as paramagnetic, while the quadratic ones as diamagnetic. The appearance of the angular momentum operator in general leads to complex wavefunctions for molecules in magnetic fields. In order to obtain gauge-origin independent observables for approximate wavefunctions, complex London orbitals\cite{London1937} can be employed
\begin{equation}
    \omega(\bm{B},\bm{O},\bm{A},\bm{r}^{\bm{O}}_i) = e^{\frac{\tn{i}}{2}\left[\bm{B}\times\left( \bm{O}-\bm{A} \right)\right]\cdot \bm{r}^{\bm{O}}_i}\chi(\bm{A},\bm{r}^{\bm{O}}_i),
\end{equation}
where $\bm{A}$ are the coordinates of the atomic center of the basis function $\chi(\bm{A,\bm{r}^{\bm{O}}_i})$. 

\subsection{Coupled-Cluster Theory}
In CC theory,\cite{Cizek1966,Shavitt2009} the electronic wavefunction is expressed as an exponential expansion of the cluster operator $\hat{T}=\sum\limits_{\mu}t_{\mu}~\hat{\Omega}_{\mu}$ acting on a reference determinant
\begin{align}
    \ket{\Psi_{\tn{CC}}}=e^{\hat{T}} \ket{\Phi_0}. \label{eq:CC}
\end{align}
$\hat{\Omega}_{\mu}$ are strings of quasiparticle creation operators $\{\hat{a}_a^{\dagger}\}$ and $\{\hat{a}_i\}$. In the notation used, $i,~j,~k,\dots$ denote occupied  and $a,~b,~c,\dots$ virtual orbitals. 
Substituting eq.~\ref{eq:CC} into the Schrödinger equation and further multiplying with $e^{-\hat{T}}$ from the left results in
\begin{equation}
    \widetilde{H}\ket{\Phi_0}=E_{\tn{CC}}\ket{\Phi_0},
\end{equation}
with $\widetilde{H}=e^{-\hat{T}}\hat{H}e^{\hat{T}}$
the similarity-transformed Hamiltonian.
The correlated energy and the cluster amplitudes are determined using projections
\begin{align}
    \bra{\Phi_\mu}\widetilde{H}\ket{\Phi_0}=\delta_{\mu0}E_{\tn{CC}},
\end{align}
onto the reference and $\mu$-level excited determinant $\Phi_\mu$.

\subsection{Equation-of-Motion ansatz}
In Equation-of-Motion CC theory,\cite{Monkhorst1977,Koch1990,Shavitt2009} the wavefunction is expressed as an operator $\hat{\mathcal{R}}$ acting on a CC reference-state wavefunction
\begin{align}
    \ket{\Psi_{\tn{EOM}}}&=\hat{\mathcal{R}}\ket{\Psi_{\tn{CC}}},
\end{align}
with $\hat{\mathcal{R}}=\sum\limits_{\mu}r_{\mu}~\hat{\Omega}_{\mu}$.
The weighting factors $r_{\mu}$ are the EOM amplitudes. 

In contrast to the standard EE-EOM formulation that preserves both the particle number and the overall spin, in SF-EOM, quasiparticle strings that change the overall spin by one ($\Delta M_S=\pm 1$) are employed.\cite{Levchenko2004,Krylov2006} Additionally in the IP and EA variants of EOM, the $\hat{\Omega}_{\mu}$ operators do not preserve the particle number resulting in an odd number of elements in the operator string:\cite{Stanton1994,Nooijen1995}
\begin{align}
    \hat{\mathcal{R}}_{\tn{IP}}&=\sum\limits_{i}r_{i}~\hat{a}_{i}
+\frac{1}{2}\sum\limits_{i,j}\sum\limits_{b}r_{ji}^{b}~\hat{a}_{b}^\dagger\hat{a}_{j}\hat{a}_{i}+\dots~, \label{R_IP} \\ 
    \hat{\mathcal{R}}_{\tn{EA}}&=\sum\limits_ar^a~\hat{a}_a^\dagger
    +\frac{1}{2}\sum\limits_j\sum\limits_{a,b}r_{j}^{ba}~\hat{a}_a^\dagger\hat{a}_b^\dagger\hat{a}_j+\dots \label{R_EA}
\end{align}
Similarly to the SF variant, the spin of the leaving or the attached electron gives rise to a change of the total $M_S$ quantum number by $\Delta M_S = \pm \frac{1}{2}$.
The use of the different EOM-CC variants is advantageous, as it allows the treatment of open-shell systems, or states that are dominated by multiple determinants starting from a well-behaved reference state. Depending on the choice of CC reference, this may eliminate spin-contamination or allow the calculation of multiconfigurational states.\cite{Krylov2017,Krylov2008}

The EOM amplitudes are found by solving the energy eigenvalue problem:\cite{Shavitt2009}
\begin{equation}
    \bra{\Phi_\mu}\widetilde{H}\hat{\mathcal{R}}\ket{\Phi_0}= E_{\tn{EOM}} r_\mu,
\end{equation}
with $E_{\tn{EOM}}$ the energy of the EOM state.
Since the similarity-transformed Hamiltonian is not a Hermitian operator, the left-side eigenvalue problem 
\begin{equation}
    \bra{\Phi_0}\hat{\mathcal{L}}\widetilde{H}\ket{\Phi_\mu}= l_{\mu} E_{\tn{EOM}},
\end{equation}
with $\hat{\mathcal{L}}=\sum\limits_{\mu}l_{\mu}~\hat{\Omega}^\dagger_{\mu}$ the EOM left-side deexcitation operator, is not the Hermitian conjugate of the right side. In the vicinity of conical intersections\cite{Kohn2007} and in the case of ff calculations, the non-Hermicity of the EOM-CC ansatz may give rise to unphysical complex energies. This behaviour has been investigated in Ref.~\onlinecite{Thomas2021}.  

The left-hand side EOM-CC problem needs to be solved only in the case of property calculations but not for energies. The left- and right-hand side operators obey the biorthonormality condition 
\begin{equation}
    \bra{\Phi_0}\hat{\mathcal{L}}^{(n)}\hat{\mathcal{R}}^{(m)}\ket{\Phi_0}=\delta_{nm}.
\end{equation}
Indices $m$,$n$ in the equation above enumerate the excited-state solutions of the eigenvalue problem. 
Within EOM-CC theory, a property described by operator $\hat{A}$ may be calculated as an biorthogonal expectation value\cite{Stanton1993,Hampe2019}
\begin{equation}
\braket{\hat{A}_{nm}}=\bra{\Phi_0}\hat{\mathcal{L}}^{(n)}\widetilde{A}\hat{\mathcal{R}}^{(m)}\ket{\Phi_0},\label{eq:EOMexp}
\end{equation}
where $\widetilde{A}=e^{-\hat{T}}\hat{A}e^{\hat{T}}$ is the similarity-transformed operator for the property of interest. 

Explicit expressions for solving the right- and left-hand side for the ff EOM-CCSD truncation as well as for the calculation of properties are given in Ref.~\onlinecite{Hampethes}.

An important difference between the field-free case and ff calculations is that the IP/EA variants do not account for the energy of the ejected/captured electron. In the presence of a magnetic field, the energy of the free electron is quantized by the Landau levels
\begin{equation}
    E^\tn{Landau}_{n,m_l,m_s} = \left(n+\frac{m_l}{2}+m_s+\frac{|m_l|}{2}+\frac{1}{2}\right)|\bm{B}|. \label{Landaulevel}
\end{equation}
This energy needs to be accounted for when IPs or EAs are calculated in the presence of a magnetic field.\cite{Holzer2019}

\subsection{The CCSD(T)(a) and EOM-CCSD(T)(a)* approach}
The CCSD(T)(a)* approach developed by \citeauthor{Matthews2016}\cite{Matthews2016} functions similarly to CCSD(T), meaning it offers perturbative triples corrections using non-iterative $M^7$ steps on top of a CCSD calculation. In contrast to CCSD(T) however, it is able to treat both ground and excited states at the CC and EOM-CC levels of theory, respectively. The method is recapitulated in the following paragraphs. 

The first step of the CCSD(T)(a)* method is to correct the CC reference-state energy and wavefunction after a CCSD calculation. Triple amplitudes are defined at the second order of perturbation using the converged CCSD amplitudes ($t^\tn{CCSD}$) 
\begin{equation}
    t_3^{[2]}= - \frac{\bra{\Phi_3}\left[\hat{V}_\tn{N},\hat{T}^\tn{CCSD}_2\right]\ket{\Phi_0}}{\Delta \varepsilon_3},
\end{equation}
with the index $\tn{N}$ denoting the normal ordering of the two-electron interaction operator $\hat{V}$ and $\Delta \varepsilon_\mu=\varepsilon_a+\varepsilon_b+\varepsilon_c + \dots -\varepsilon_i-\varepsilon_j-\varepsilon_k - \dots $ the orbital energy difference between determinants $\Phi_0$ and $\Phi_\mu$. The calculation of the pertubative triple amplitudes  is an $M^7$ step.  Using $\hat{T}_3^{[2]}$, the converged CCSD amplitudes are corrected
\begin{align}
    t_1^\tn{corr}&=t_1^\tn{CCSD} - \frac{\bra{\Phi_1}\left[\hat{V}_\tn{N},\hat{T}^{[2]}_3\right]\ket{\Phi_0} }{\Delta \varepsilon_1}\\
    t_2^\tn{corr}&=t_2^\tn{CCSD} - \frac{\bra{\Phi_2}\left[\hat{F}_\tn{N} + \hat{V}_\tn{N},\hat{T}^{[2]}_3\right]\ket{\Phi_0} }{\Delta \varepsilon_2},
\end{align}
where $\hat{F}_\tn{N}$ is the normal-ordered Fock operator. Using the corrected amplitudes, the CCSD(T)(a) energy is given by
\begin{equation}
    E_\tn{CCSD(T)(a)}=\bra{\Phi_0} e^{-\hat{T}^\tn{corr}} \hat{H} e^{\hat{T}^\tn{corr}} \ket{\Phi_0} = \bra{\Phi_0}  \widetilde{H}^\tn{corr}  \ket{\Phi_0}.
\end{equation}

For the EOM-CC part, two kinds of triples corrections, an implicit and an explicit one are employed. The implicit correction is performed by using the corrected $t^\tn{corr}_\mu$ amplitudes. This leads to the EOM-CCSD(T)(a)$^0$ eigenvalue problem which takes the form
\begin{align}
    \bra{\Phi_\mu}\widetilde{H}^\tn{corr} \left(\hat{\mathcal{R}}_1+\hat{\mathcal{R}}_2\right) \ket{\Phi_0} &= E_{\tn{EOM-CCSD(T)(a)}^0} r_\mu \\
\end{align}
and 
\begin{align}
    \bra{\Phi_0}\left(\hat{\mathcal{L}}_1+\hat{\mathcal{L}}_2\right) \widetilde{H}^\tn{corr}  \ket{\Phi_\mu} &= l_\mu E_{\tn{EOM-CCSD(T)(a)}^0} ,
\end{align}
for $\mu=0,1,2$.
Important to note is that the non-vanishing overlaps $\bra{\Phi_\mu} \widetilde{H}^\tn{corr}  \ket{\Phi_0}$ have been deliberately projected out to retain size-consistency and preserve the $M^6$ scaling of EOM-CCSD. Solving the EOM-CCSD(T)(a)$^0$ problem uses the exact same routines as EOM-CCSD. 

Beyond the implicit triples contributions to the excitation energy discussed so far, the EOM-CCSD* approach is used to account for direct triples contributions from the EOM vectors.\cite{Stanton1996} Triples EOM vectors $\hat{\mathcal{L}}^*_3$ and $\hat{\mathcal{R}}^*_3$ are defined in the first non-vanishing order of perturbation. Their calculation scales as $M^7$ for the EE and SF variants and as $M^6$ for IP and EA. The final EOM-CCSD(T)(a)* energy takes the form
\begin{equation}
    E_{\tn{EOM-CCSD(T)(a)}^*}=E_{\tn{EOM-CCSD(T)(a)}^0} + \bra{\Phi_0}\frac{\hat{\mathcal{L}}^*_3\hat{\mathcal{R}}^*_3}{\Delta E_{\tn{EOM-CCSD(T)(a)}^0} - \Delta \varepsilon_3}\ket{\Phi_0},
\end{equation}
where $\Delta E_{\tn{EOM-CCSD(T)(a)}^0}=E_{\tn{EOM-CCSD(T)(a)}^0}-E_{\tn{CCSD(T)(a)}}$ is the excitation energy at the EOM-CCSD(T)(a)$^0$ level of theory. 
Explicit expressions and working equations for the ff implementation of the different EOM variants can be found in Ref.~\onlinecite{Kitsarasthes}. Important to note is that this approach is designed to target states with a single-excitation character with respect to the reference state and is inappropriate when double-excitation character is dominant.

%%%%%%%%%%%%%%%%%%%%%%%%%%%%%%%%%%%%%%%%%%%%%%%%%%%%%%%%%%%%%%%%%%%%%%%%%%%%
\section{\label{Implementation}Implementation}
%%%%%%%%%%%%%%%%%%%%%%%%%%%%%%%%%%%%%%%%%%%%%%%%%%%%%%%%%%%%%%%%%%%%%%%%%%%%
The ff complex-valued SF/IP/EA-EOM-CCSD methods have been implemented within the QCUMBRE program package\cite{qcumbre,Hampe2017} including the calculation of one-electron properties following the expectation-value approach as presented in Refs.~\onlinecite{Stanton1993,Hampe2019}. Moreover, approximate triples at the CCSD(T)(a) and EOM-CCSD(T)(a)* level have been implemented in the program for the EE, SF, IP, and EA variants. 

In the case of the ff-SF variant, the code was tested against the ff-EE implementation, i.e., it was checked that the same result is obtained by calculating the $M_S=0$ triplet within EE-EOM, and the $M_S=\pm 1$ triplet using SF-EOM while disregarding the spin-Zeeman contribution. For the IP/EA-EOM implementation, results were validated against EE results that make use of continuum orbitals to model the electron ejection/capture, respectively.\cite{Stanton1999} 
As for the implementation of the CCSD(T)(a) and  EE-EOM-CCSD(T)(a)* approach, the implementation was validated against the CFOUR\cite{Matthews2020,cfour} implementation in the field-free case. Further details on the validation can be found in Refs.~\onlinecite{Hampethes,Kitsarasthes}.

%%%%%%%%%%%%%%%%%%%%%%%%%%%%%%%%%%%%%%%%%%%%%%%%%%%%%%%%%%%%%%%%%%%%%%%%%%%%
\section{\label{Applications}Applications}
%%%%%%%%%%%%%%%%%%%%%%%%%%%%%%%%%%%%%%%%%%%%%%%%%%%%%%%%%%%%%%%%%%%%%%%%%%%%

All calculations have been performed using the Hartree-Fock solver either  in the LONDON\cite{LONDON} or in the CFOUR\cite{Matthews2020,cfour} program interfaced to the QCUMBRE\cite{qcumbre} program package. 
When using CFOUR, the required integrals over London orbitals,\cite{London1937} which have been employed in all calculations, are provided by means of the MINT integral package.\cite{MINT}   
Throughout this paper, electronic states are labeled as A/B. A refers to the field-free irreducible representation (IRREP)  and B is the respective IRREP in the presence of the magnetic field.
For simplicity, when listing occupations, we limit ourselves to the field-free notation. For open-shell states, the component with the lowest possible $M_S$ value is calculated, unless stated otherwise. 

We study the ionization potentials and electron affinities of the first 10 main group elements in the presence of an increasingly strong magnetic field. As pointed out in previous studies\cite{Holzer2019} and will be explained in the next section, a perturbative consideration predicts no field-dependence for the IPs and EAs, as the paramagnetic contributions cancel out between the ejected or captured electron and the non-ionized species (see \eqref{eq:IEDef2} and \eqref{eq:EADef1}). The diamagnetic contributions, however, cause significant alterations from the perturbative predictions. The latter are considered when using the ff methodology and play an important role for strong fields as occurring on magnetic WDs. 
Furthermore, calculations were performed to study the behaviour of heavier elements Na, Mg and Ca in strong magnetic fields. Specifically, the influence of triples corrections to the IPs of Na and Mg, and to the electronic excitation between triplet states of Ca was considered. The results complement the investigation of the relevant transitions of these atoms, which have been detected on a strongly magnetic WD star.\cite{Hollands2023} 
Lastly, the behaviour of the low-lying states of the CH molecule was explored in  a strong magnetic field. Since the molecule has been detected in in weakly magnetic WDs,\cite{Berdyugina2007,Vornanen2010,Vornanen2013} one can expect that it occurs in strongly magnetic WDs as well.

\subsection{Ionization potentials and electron affinities in strong magnetic fields}
In order to better understand the conditions in the atmospheres of magnetic WDs, we study the IPs and EAs using the ff IP- and EOM-CC approaches. As the names of the methods imply, they can be directly applied to such investigations, giving access to a whole set of ionized states within a single calculation. \citeauthor{Nooijen1995} were the first to present the EA-EOM approach and compared its performance to calculate the EAs to two consecutive CCSD calculations for the neutral and ionic system, i.e., the $\Delta$CCSD method. Their findings suggest that EA-EOM performs very similarly to $\Delta$CCSD but with a reduced computational cost. A comparison of the performance of $\Delta$CCSD versus IP-EOM  and EA-EOM in an increasingly strong magnetic field for the Li atom seems to validated this claim also for strong magnetic fields. The respective calculations are presented in section I of the SI.

\subsubsection{Ionization potentials \label{sssec:IP}}

The IPs of the first ten elements have been  calculated in Ref.~\onlinecite{Holzer2019} as a function of the magnetic-field strength using Green's-functions techniques and were benchmarked against IP-EOM-CCSD data. The main discussion points will be reiterated for a better understanding of the similarities and differences between IPs and EAs (see next chapter) in a strong magnetic field.

Calculations were carried out for $B\leq 0.25\BB$ using the same basis set as in the previous study, i.e., the Cartesian uncontracted doubly augmented (cart-unc-d-aug) cc-pwCVQZ basis.\cite{Holzer2019}  
The first IP was determined using the following protocol: First, the ground state of the neutral atom was determined at the CCSD level. Then, two sets of IP-EOM-CCSD calculations ($\Delta M_S=+\frac{1}{2}$ and $\Delta M_S=-\frac{1}{2}$) were performed using that CCSD reference wave function to determine the lowest IP.  

\begin{figure}[!htb]
\centering

\includegraphics[width=10cm]{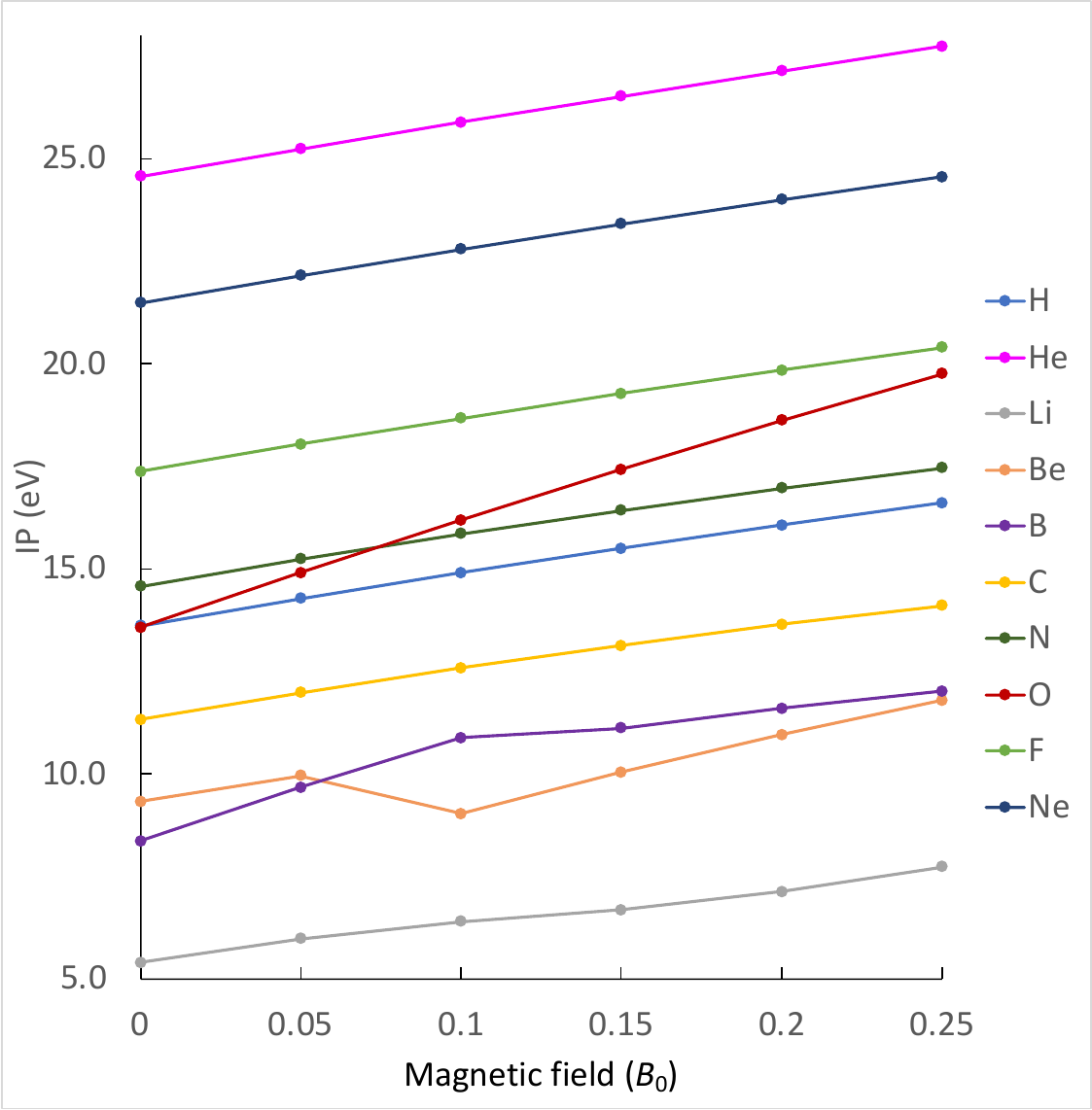} 
\caption{Landau-corrected IPs of the atoms H-Mg  in a magnetic field between $0\tn{-}0.25\BB$ .}
\label{fig:IPs}
\end{figure}
As discussed in Ref.~\onlinecite{Holzer2019}, the interpretation of IP-EOM energies as ionization energies is correct in the field-free case but the same does not apply in the presence of a magnetic field. 
This  is due to the fact that the energy of the ionized electron is nonzero. 
However, this contribution is not considered within the EOM framework. Instead, the electron is removed entirely from the system. 
The IP for the ionization process 
\begin{equation*}
\tn{M} \rightarrow \tn{M}^+ + e^-
\end{equation*}
generating a state $\mathrm{S2}$ from a state $\mathrm{S1}$ must, therefore, be calculated as
\begin{equation}
\mathrm{IP}=E^\mathrm{S2} -E^\mathrm{S1}+E^\mathrm{L} , 
\label{eq:IEDef1}
    \end{equation}
with the lowest Landau energy $E^\mathrm{L}$, see \eqref{Landaulevel}.
Taking the corresponding corrections $E^L$ into account, the IPs depicted in \figref{fig:IPs} are obtained for the atoms H-Ne. The corresponding ionization channels can be found in Ref.~\onlinecite{Holzer2019}.  
The most striking observation is that for the considered range, all IPs increase with the field strength in a concave manner. 
Decomposing the paramagnetic and diamagnetic contributions in \eqref{eq:IEDef1} results in
\begin{align}
\notag \tn{IP}&=E^\mathrm{S2}-E^\mathrm{S1}+E^\mathrm{L}\\
\notag &= \Delta E_0^\mathrm{S2-S1}+\Delta E_\text{para}^\mathrm{S2-S1}+\Delta E_\text{dia}^\mathrm{S2-S1}+\underbrace{\left(\frac{1}{2}m_l+m_s\right)
\cdot B}_{E^\mathrm{L}_\text{para}}+\underbrace{\frac{1}{2}\left(1+\lvert m_l\rvert\right)\cdot B}_{E^\mathrm{L}_\text{dia}}\\
\notag&=\Delta E_0^{S2-S1}+\Delta E_\text{dia}^{S2-S1}+E_\text{dia}^\mathrm{L}+\underbrace{\Delta E_\text{para}^\mathrm{S2-S1}+E_\text{para}^\mathrm{L}}_{=0^*}\\
\label{eq:IEDef2}&=\underbrace{\Delta E_0^\mathrm{S2-S1}}_{\mathcal{O}(B^0)}+\underbrace{E_\text{dia}^\mathrm{L}}_{\mathcal{O}(B^1)}+\underbrace{\Delta E^\mathrm{S2-S1}_\text{dia}}_{\mathcal{O}(B^2)}.\\
\notag^*&{\text{only if the total angular momentum quantum number}M_L\text{ is a good quantum number}}
\end{align}
$\Delta E_0^\mathrm{S2-S1}$ designates the field-independent energy difference between the initial and final state and does not scale with $B$. However, it is not constant for different magnetic-field strengths because the orbitals as well as the corresponding amplitudes are optimized for a given $B$. The term also defines the origin of all ionization-energy curves at $B=0 \BB$. As mentioned in Ref.~\onlinecite{Holzer2019}, the Landau energy is comprised of a paramagnetic and a diamagnetic contribution. In \eqref{eq:IEDef2}, all paramagnetic contributions cancel out exactly.   
The diamagnetic part $\left(1+\lvert m_l\rvert\ \right) \cdot \frac{B}{2}$ of the Landau energy is linear in $B$ and always has a positive contribution. This term  defines the initial slope which is always positive. For the systems studied, two different initial slopes are observed which reflect the orbital character of the ejected electron. The latter is ejected either from an $s$ or a $p$ orbital. For $s$ and $p_0$ orbitals ($m_l=0$), the initial slope is $\frac{1}{2}\EhB $ and for ionizations from  $p_{\pm 1}$ orbitals, it  is $1\EhB  $.
The difference in the diamagnetic contribution for the states $\mathrm{S1}$ and $\mathrm{S2}$, i.e., $\Delta E^\mathrm{S2-S1}_\text{dia}$, scales with $B^2$. As the diamagnetic contribution for the $(N-1)$-electron system is expected to be smaller as compared to the corresponding $N$-electron system, this term has a negative sign. Hence, the magnitude of the initial slope decreases with increasing field strength and the observed concave functions are obtained. In the field range considered here, ionization is less favorable with increasing field strength as compared to the field-free case. For stronger magnetic fields, however, the diamagnetic contribution is expected to become larger and ionization will become easier with increasing the magnetic-field strength. 

It should be emphasized that the ionization path does not necessarily stem from the highest occupied molecular orbital (HOMO). This is due to correlation and relaxation effects and because of the quantized Landau levels of the free electron. A further discussion for the case of Carbon can be found in section II of the SI. In the case of Neon, ionization arises from ejection from the  $2p_0$ orbital rather than from the $2p_{+1}$ orbital the latter of which is the HOMO for all non-zero field strengths considered. In fact, as the paramagnetic contributions have no effect on the IPs, for different channels that share $\Delta E_0^{S2-S1}$, the spin and the angular momentum of the leaving electron do not contribute to the IP.  
The second striking feature in \figref{fig:IPs} concerns the discontinuities in the IP curves for Li, B, and, Be. They stem from a change of the ground state for the respective atoms at a certain magnetic-field strength. For Li, the different IPs are very similar making the discontinuity less pronounced. 
For even higher field strengths, such discontinuities in the evolution of the IPs due to a change in the preferred ionization process are to be expected. For example, while the IP for ionizing from the $2p_0$ orbital in the oxygen atom is higher than for the $2p_{-1}$ orbital in the field-free case, the slope of the IP for the former process ($\frac{1}{2}\EhB $) is smaller compared to the one for the latter ($1\EhB  $). Hence, the respective IP curves will eventually cross and ionization from the $2p_0$ orbital will become energetically more favorable.

Concluding, the evolution of IPs for the first and second row atoms as a function of the magnetic field is essentially governed by the diamagnetic contribution of the energy of the ejected electron.\cite{Holzer2019} It increases the energy necessary to ionize an atom in a magnetic field.   An eventual decrease is expected in stronger fields.

\subsubsection{Electron affinities \label{sssec:EA}}

EAs were calculated for the first ten elements of the periodic table at the EA-EOM-CCSD level of theory using the cart-unc-d-aug-pwCVQZ basis set for a magnetic field up to $0.25\BB$. A similar protocol as for the calculation of the IPs was applied, i.e., EA-EOM-CCSD energies were calculated using the respective ground state of the neutral system for each magnetic-field strength as a reference at the CCSD level of theory in order to find the most favorable EA path. Similar considerations as for the IPs regarding the Landau energy of the captured electron need to be applied. The EA, for which 
\begin{equation*}
    \tn{M} + e^- \rightarrow \tn{M}^-
\end{equation*}
describes the energy difference between an $(N+1)$-system in state $\mathrm{S2}$ and the preceding $N$-electron system in state $\mathrm{S1}$ and is hence given by
\begin{equation}
    \mathrm{EA}=E^\mathrm{S2}-E^\mathrm{S1}-E^\mathrm{L}. \label{eq:EADef1}
\end{equation}

\begin{figure}[!htb]
    \centering
    \includegraphics[width=10cm]{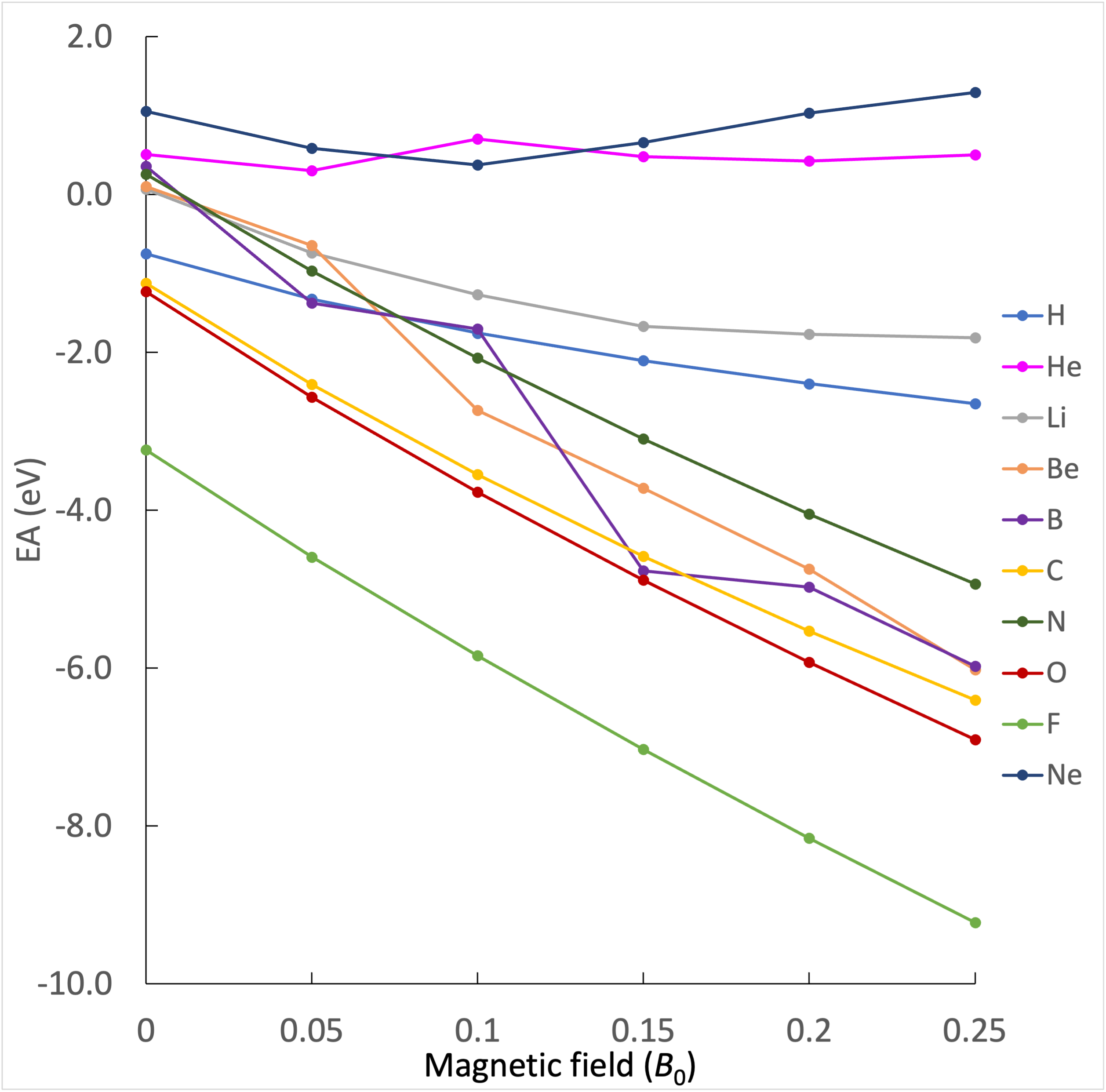}
    \caption{Landau-corrected EAs of the atoms H-Ne in a magnetic field between 0-0.25 B$_0$.}
\label{fig:EAs}
\end{figure}

The physical EAs, i.e., those that include the Landau energy, are shown in \figref{fig:EAs} for the elements H-Ne. 
At a first glance, the EAs seem to evolve less smoothly with increasing magnetic-field strength compared to the corresponding IPs in Fig.~\ref{fig:IPs}. Most EAs are decreasing while some increase parabolically for stronger magnetic-field strengths. All curves exhibit concave behavior which is explained by rewriting \eqref{eq:EADef1} in an analogous way as for the IPs (see eq.~\ref{eq:IEDef2})
\begin{equation}
    \mathrm{EA}=\underbrace{\Delta E_0^\mathrm{S2-S1}}_{\mathcal{O}(B^0)}-\underbrace{E_\text{dia}^\mathrm{L}}_{\mathcal{O}(B^1)}+\underbrace{\Delta E^\mathrm{S2-S1}_\text{dia}}_{\mathcal{O}(B^2)}. \label{EADef2}
\end{equation}
The initial slope is always negative as the Landau energy has to be subtracted in order to obtain EAs. The exact value of the slope is given by the Landau energy term and is again dictated by the azimuthian quantum number of the attached electron, resulting in $-\frac{1}{2}\EhB$ for $m_l=0$ and $-1\EhB  $ for $|m_l|=1$. 
However, the ${\Delta E_\text{dia}^\mathrm{S2-S1}}$ contribution which depends quadratically on $B$ is positive for the electron attachment. This leads  to the concave behaviour of the EA as a function of the magnetic-field strength. 
As the Landau energy is subtracted in \eqref{EADef2}, capturing electrons with a large $\lvert m_l\rvert$ is energetically favorable. 
For the nitrogen atom for example, the EA channel that involves the $2p_{\pm 1}$ orbital is more beneficial compared to the $2p_0$ orbital. 
This and the fact that electron spin has no influence since the paramagnetic contribution cancels out means that  the most favorable attachment process does not necessarily produce the ground state of the anionic system. E.g., in the case of boron, the $^2P_u/^2\Pi^-_u$ ground state of the neutral system with an electron configuration $1s^2 2s^2 2p_{-1}$  preferably captures a $2p_{+1}$ rather than a $2p_0$ electron, even though the $^3P_g/^3\Pi^-_g$ ($1s^2 2s^2 2p_{-1} 2p_{0}$) state is energetically lower than $^3P_g/^3\Sigma_g$ ($1s^2 2s^2 2p_{-1}2p_{+1}$) for the anion due to the paramagnetic stabilization. For a more detailed discussion of the Landau contribution see section II in the SI where the energies of different states of the fluoride anion are compared. 

In contrast to the IPs, for the EAs the diamagnetic contribution $\Delta E^\mathrm{S2-S1}_\text{dia}$ has a larger influence in the considered range of field strengths. This is most apparent for the noble gas atoms He and Ne for which the additional electron occupies an orbital in a higher shell making the system even more diffuse. The behaviour of the EAs of the atoms H, N, O, and F in the magnetic field strengths studies here exhibits no discontinuities and no change of the ground state of the system. Discontinuities in the curves for Li, Be, and B are traced back to changes in the electronic ground state, similar to the discussion of the IPs. Abrupt change of the slope signifies a change of the electron capturing path as can be observed in the case of He, C, and Ne. The most favourable capturing channels for different magnetic-field strengths are summarized in Tab.~\ref{tab:EAs}.

\begin{table}[]
    \centering
    \begin{tabular}{ccc}
        Element & Ground State & Orbital of attached electron \\ \hline \hline
        H & $^2S_g / ^2\Sigma_g$ ($1s^\beta$) & $1s^\alpha$ \\ \hline
        He & $^1S_g/ ^1\Sigma_g$ ($1s^2$) & $2s$, $B < 0.106\BB$ \\ 
        ~ & ~ & $2p$, $B > 0.106\BB$ \\ \hline
        Li & $^2S_g/ ^2\Sigma_g$ ($1s^2 2s^\beta$), $B < 0.175\BB$ & $2s^\alpha$, $B < 0.090\BB$ \\ 
        ~ & ~ & $2p^\beta$, $B > 0.090\BB$ \\ 
        ~ & $^2P_u/ ^2\Pi^-_u$ ($1s^2 2p_{-1}^\beta$), $B>0.175\BB$ & $2s^\beta$, $B < 0.210\BB$ \\
        ~ & ~ & $2p_{-1}^\alpha$, $B > 0.210\BB$ \\ \hline
        Be & $^1S_g/ ^1\Sigma_g$ ($1s^2 2s^2$, $B < 0.067\BB$ & $2p$ \\ 
        ~ & $^3P_u/ ^3\Pi^-_u$ ($1s^2 2s^\beta2p_{-1}^\beta$), $B > 0.067\BB$  & $2s^\alpha$, $B  < 0.149\BB$ \\ 
        ~ & ~ & $2p_{+1}^\beta$, $B  > 0.149\BB$ \\ \hline
        B & $^2P_u/ ^2\Pi^-_u$ ($1s^2 2s^2 2p_{-1}^\beta$), $B < 0.126\BB$ & $2p_{+1}^\beta$ \\ 
        ~ & $^4P_g/ ^4\Pi^-_g$ ($1s^2 2s^\beta 2p_{-1}^\beta 2p_{0}^\beta$), $B > 0.126\BB$ & $2s^\alpha$, $B  > 0.193\BB$ \\ 
        ~ & ~ & $2p_{+1}^\beta$, $B  > 0.193\BB$ \\ \hline
        C & $^3P_g/ ^3\Pi^-_g$ ($1s^2 2s^2 2p_{-1}^\beta 2p_{0}^\beta$)  & $2p_{+1}^\beta$ \\ \hline
        N & $^4S_u/ ^4\Sigma_u$ ($1s^2 2s^2 2p_{-1}^\beta 2p_{0}^\beta 2p_{+1}^\beta$)  & $2p_{\pm1}^\alpha$ \\ \hline
        O & $^3P_g/ ^3\Pi^-_g$ ($1s^2 2s^2 2p_{-1}^2 2p_{0}^\beta 2p_{+1}^\beta$)  & $2p_{+1}^\alpha$ \\ \hline
        F & $^2P_u/ ^2\Pi^-_u$ ($1s^2 2s^2 2p_{-1}^2 2p_{0}^2 2p_{+1}^\beta$)  & $2p_{+1}^\alpha$ \\ \hline
        Ne & $^1S_g/ ^1\Sigma_g$ ($1s^2 2s^2 2p^6$) & 3s, $B  < 0.045\BB$ \\ 
        ~ & ~ & 3p, $B  > 0.045\BB$ \\ 
    \end{tabular}
    \caption{Ground states and electron attachment paths for the elements of the first and second row of the periodic table up to $B=0.25 \BB$}
    \label{tab:EAs}
\end{table}

To summarize, the evolution of the EAs for the first and second row elements with an increasing magnetic-field strength is strongly dictated by the Landau energy of the captured electron. This leads to decreasing EAs with increasing magnetic field strengths. 
For weaker fields, the electron attachment process is more favorable as compared to the field-free case. 
However, the situation becomes more complex for stronger fields as the electronic diamagnetic contribution $\Delta E^\mathrm{S2-S1}_\text{dia}$ has a greater influence. This phenomenon is especially prominent when the electron attachment involves an orbital with a higher principal quantum number ($n$) as is the case for noble gas atoms. 

%%%%%%%%%%%%%%%%%%%%%%%%%%%%%%%%%%%%%%%%%%%%%%%%%%%%%%%%%%%%%%%%%%%%%%%%%%%%
\subsection{The metals Na, Mg, and Ca in a strong magnetic field}
In this section, we study the electronic structure of  Na, Mg, and Ca in the presence of strong magnetic fields. These metals are of interest for studying the atmospheres of magnetic WDs as they have been discovered in  the strongly magnetic WD SDSS J1143+6615.\cite{Hollands2023} Specifically, we investigate the IPs of Na and Mg in the presence of a magnetic field, complementing previous studies on the electronic excitations of these systems in Ref.~\onlinecite{Hampe2020} and \onlinecite{Kitsaras2024}. Moreover, the influence of the magnetic field to the electronic excitation between triplet states of Ca for transitions relevant for WD spectra is studied.  
In order to provide accurate data that may be relevant for the study of magnetic WDs, we include also the effects of triples excitations at the CCSD(T)(a)* level of theory. 

\subsubsection{Ionization potentials of Na and Mg}

\begin{figure}[!htb]
\centering
    \begin{subfigure}{0.48\textwidth}
        \begin{tikzpicture}
            \node[anchor=south west,inner sep=0] (image) at (0,0) {\includegraphics[width=\textwidth]{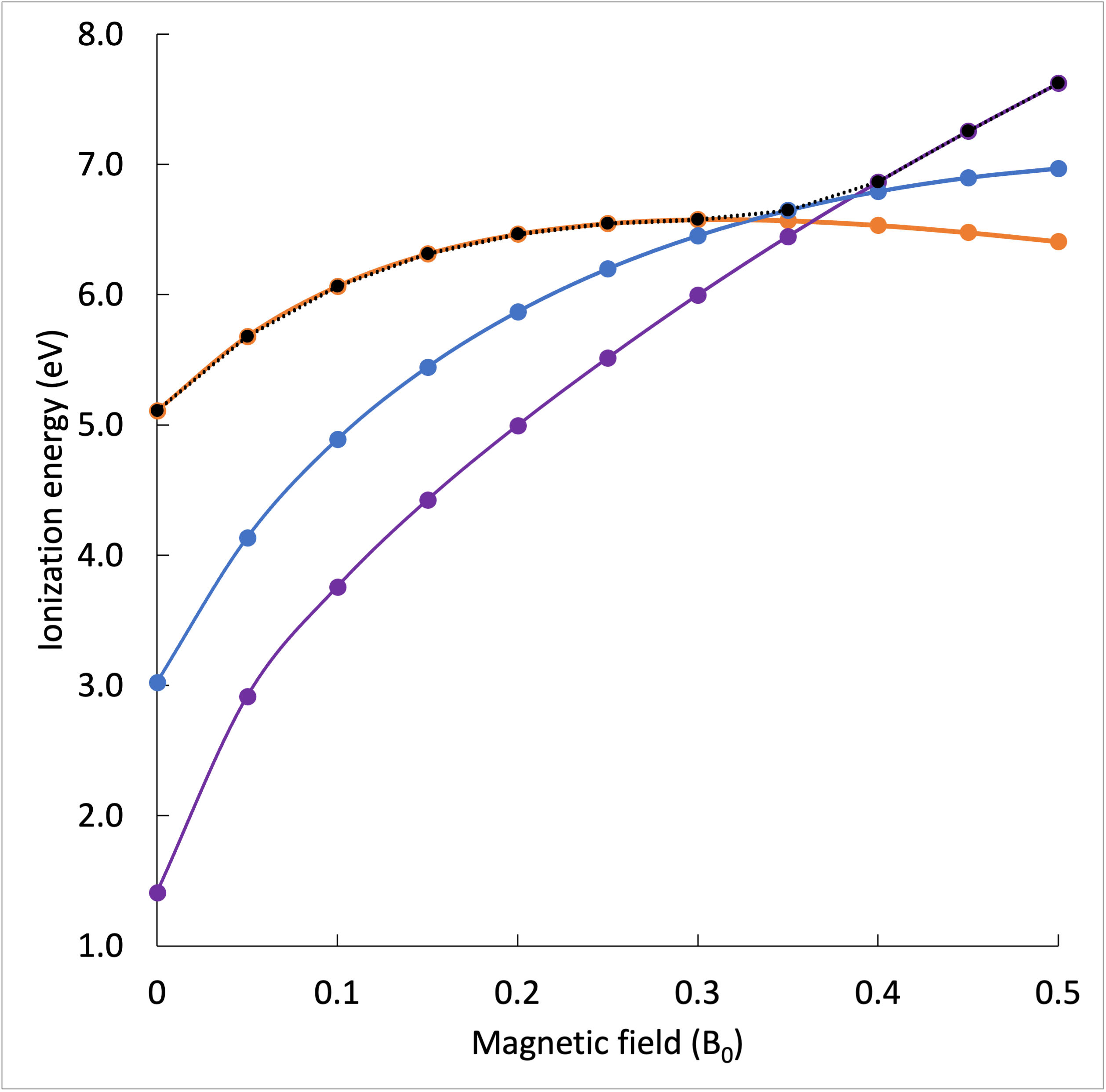}};
            \begin{scope}[x={(image.south east)},y={(image.north west)}]
                \node at (0.4,0.85) (label1) {\footnotesize $^2S_g \rightarrow {^1S}_g$};
                \node at (0.32,0.62) (label2) {\footnotesize  $^2P_u \rightarrow {^1S}_g$};
                \node at (0.7,0.65) (label3) {\footnotesize  $^2D_g \rightarrow {^1S}_g$};

                \node[anchor=west] at (0.65,0.25)  {\scalebox{0.9}{\footnotesize   CCSD}};
                \node[anchor=west] at (0.65,0.2)  {\scalebox{0.9}{\footnotesize   CCSD(T)(a)*}};
                \draw[-] (0.65,0.2) -- +(-0.06,0) node [midway] (L) {};
                \filldraw[black] (L)  circle  (1.pt);
                \draw[-] (0.65,0.25) -- +(-0.06,0);
            \end{scope}
        \end{tikzpicture}
        \caption{Na} \label{fig:IPNa}
    \end{subfigure}
    \hspace*{\fill}
    \begin{subfigure}{0.48\textwidth}
        \begin{tikzpicture}
            \node[anchor=south west,inner sep=0] (image) at (0,0) {\includegraphics[width=\textwidth]{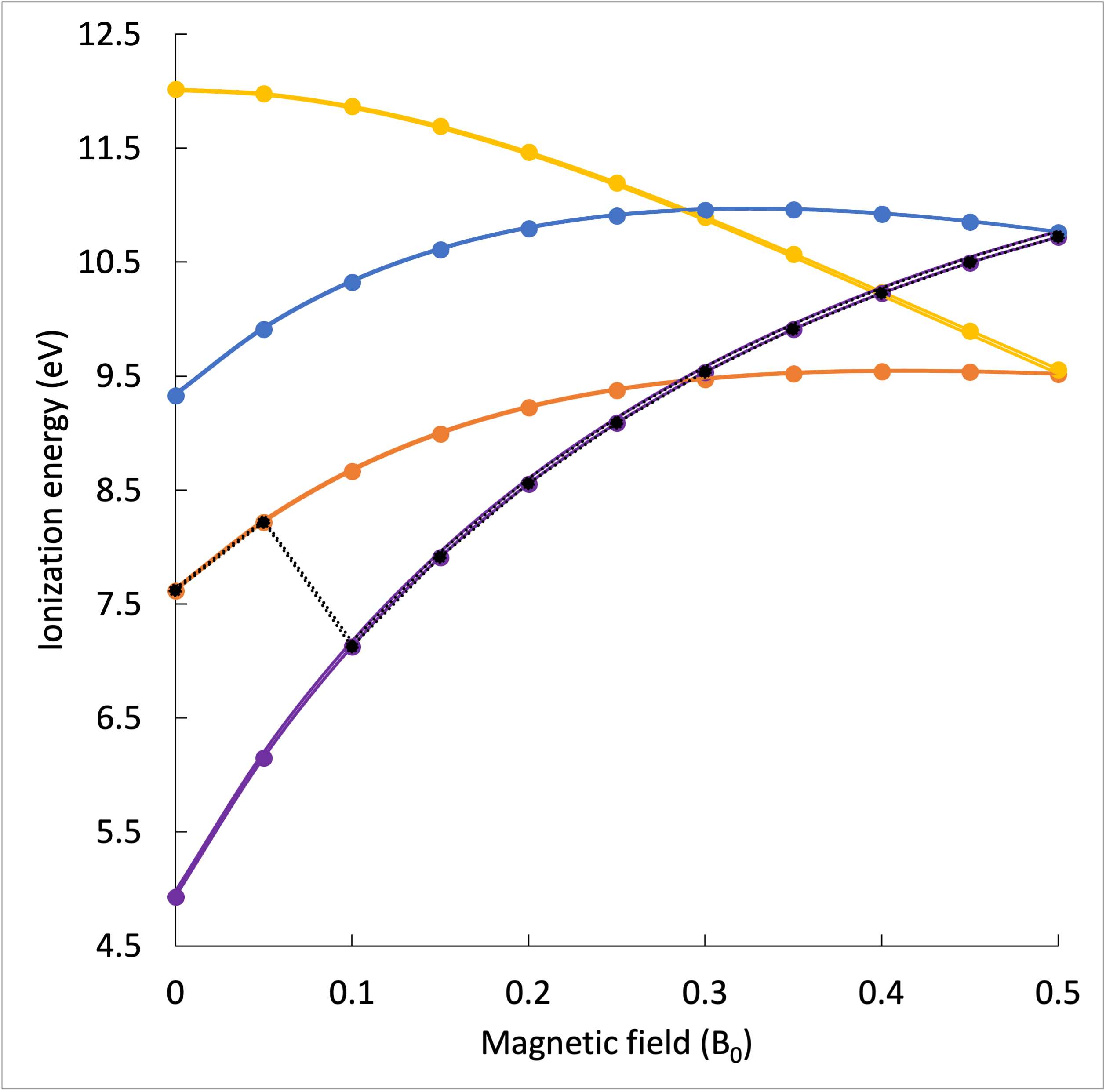}};
            \begin{scope}[x={(image.south east)},y={(image.north west)}]
                \node at (0.35,0.6) (label1) {\footnotesize $^1S_g \rightarrow {^2S}_g$};
                \node at (0.35,0.95) (label2) {\footnotesize $^1S_g \rightarrow {^2P}_u$};
                \node at (0.35,0.25) (label3) {\footnotesize $^3P_u \rightarrow {^2S}_g$};
                \node at (0.5,0.75) (label4) {\footnotesize $^3P_u \rightarrow {^2P}_u$};
                \node[anchor=west] at (0.65,0.25)  {\scalebox{0.9}{\footnotesize CCSD}};
                \node[anchor=west] at (0.65,0.2)  {\scalebox{0.9}{\footnotesize CCSD(T)(a)*}};
                \draw[-] (0.65,0.2) -- +(-0.06,0) node [midway] (L) {};
                \filldraw[black] (L)  circle  (1.pt);
                \draw[-] (0.65,0.25) -- +(-0.06,0);
            \end{scope}
        \end{tikzpicture}
        \caption{Mg} \label{fig:IPMg}
    \end{subfigure}
    
     \begin{subfigure}{0.48\textwidth}
         \begin{tikzpicture}
             \node[anchor=south west,inner sep=0] (image) at (0,0) {\includegraphics[width=\textwidth]{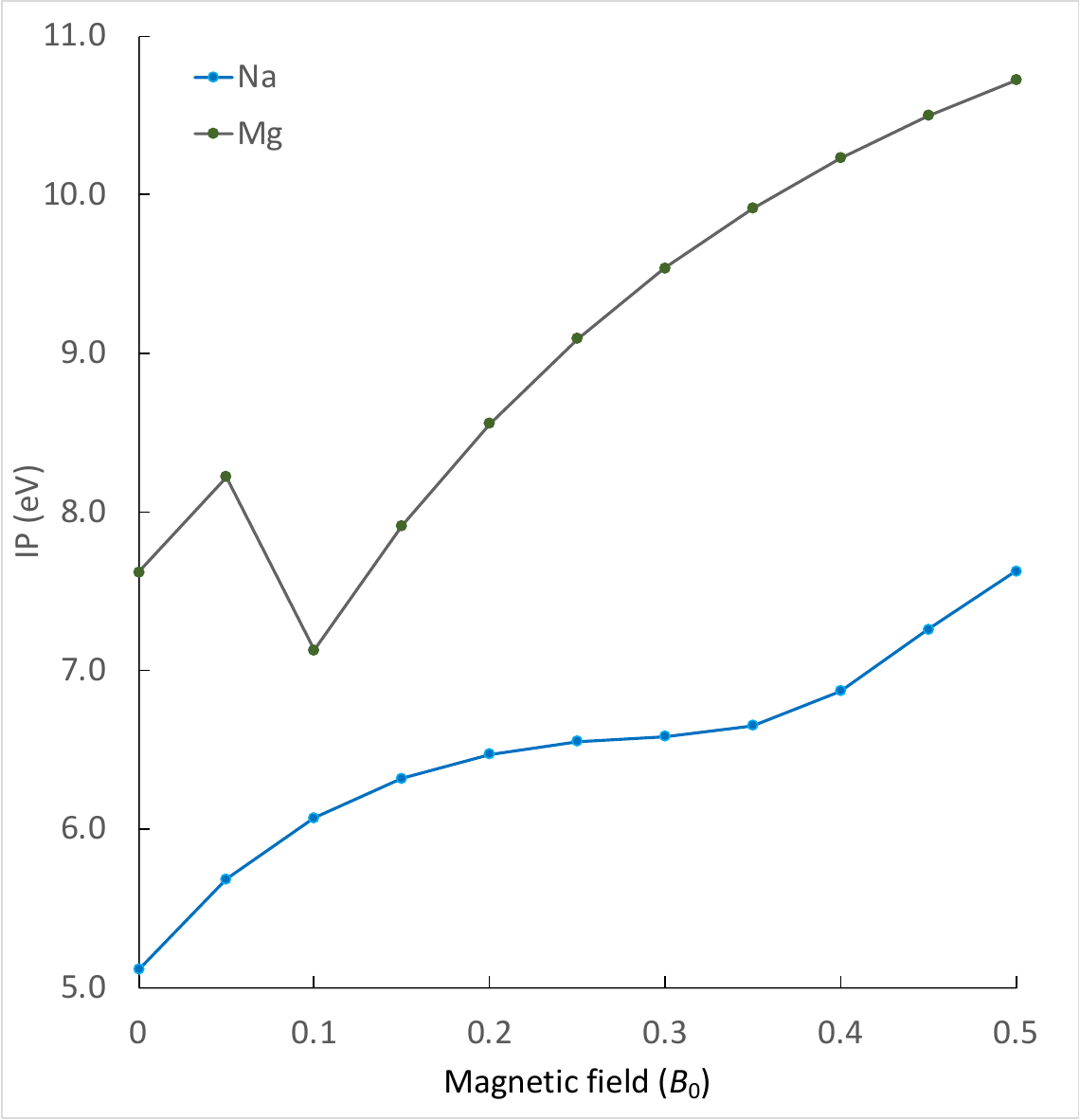}};
             \begin{scope}[x={(image.south east)},y={(image.north west)}]
             \end{scope}
         \end{tikzpicture}
         \caption{IPs} \label{fig:IPNaMg}
     \end{subfigure}
    \caption{Lowest Landau-corrected ionization paths for Na (a) and Mg (b) at the CCSD and CCSD(T)(a)* levels of theory. The corresponding lowest IPs as a function of the magnetic-field strength at the CCSD(T)(a)* level of theory (c). }
    \label{fig:astroIPs}
\end{figure}

The IPs of Na and Mg are studied  
in magnetic fields up to  $B\leq 0.50\BB$ at the EOM-CCSD and EOM-CCSD(T)(a)* levels of theory using a spherical (sph) unc-aug-cc-pVQZ basis set. For the IPs of Na, the closed-shell $^1S_g /^1\Sigma_g$ state of the cation was treated at the CC level, while the $^2S_g/^2\Sigma_g$, $^2P_u/^2\Pi_u^-$, and $^2D_g/^2\Delta_g^-$  states of the neutral atom were targeted as EA-EOM-CC states. 
For Mg, the closed-shell $^1S_g /^1\Sigma_g$ state of the neutral atom was used as a CC reference to target the $^2S_g /^2\Sigma_g$ state of the cation using the IP-EOM-CC approach. 
In addition, the $^3P_u /^3\Pi_u^-$ state of the neutral atom was treated as reference states to calculate the $^2P_u /^2\Pi_u^-$ state of the cation at the IP-EOM-CC level.

The lowest ionization paths are plotted as functions of the magnetic field in \figref{fig:astroIPs}. The first ionizations are designated with a black dotted line  and also presented separately in \figref{fig:IPNaMg}. For these elements, the divergence from linearity is more prominent as compared to the lighter elements of the first and second row. This can be traced back to the diamagnetic contribution for the electronic energy, which is more important for the larger atoms of the third period. For Na, the leaving electron is ejected from the  $3s$ orbital of the $^2S_g/^2\Sigma_g$ state for $ B = 0 \tn{-}0.3 \BB$, from the  $3p_{-1}$ orbital of the $^2P_u/^2\Pi_u^-$ state for $B=0\tn{-}0.35 \BB$, and from the  $3d_{-2}$ orbital of the $^2D_g/^2\Delta_g^-$ state for $B\geq 0.4 \BB$. For Mg, the path of the first ionization changes as well. The electron is ejected from the $3s$ orbital of the $^1S_g/^1\Sigma_g$ state for weaker fields, and from the the $3p_{-1}$ orbital of the $^3P_u/^3\Pi_u^-$ state for stronger fields. This follows the change of the ground state which occurs for $B\approx0.1\BB$. The equivalent change of ground state is observed for Be in stronger magnetic-field strengths. Triples corrections at the CCSD(T)(a)* level of theory  amount to  $\sim 0.1\mEh$ for the IPs of Na and Mg. On the scale of the plots, the difference between the CCSD and the CCSD(T)(a)* results is not visible and ranges around $\sim0.1 \mEh$. 
For a discussion of the development of the electronic states of the Na and Mg monocations see Ref.~\onlinecite{Kitsarasthes}.

\subsubsection{Electronic excitation of the Ca atom}

\begin{figure}[!htb]
\centering
\begin{minipage}[t]{0.48\textwidth}
    \begin{tikzpicture}
    \node[anchor=south west,inner sep=0] (image) at (0,0) {\includegraphics[width=\textwidth]{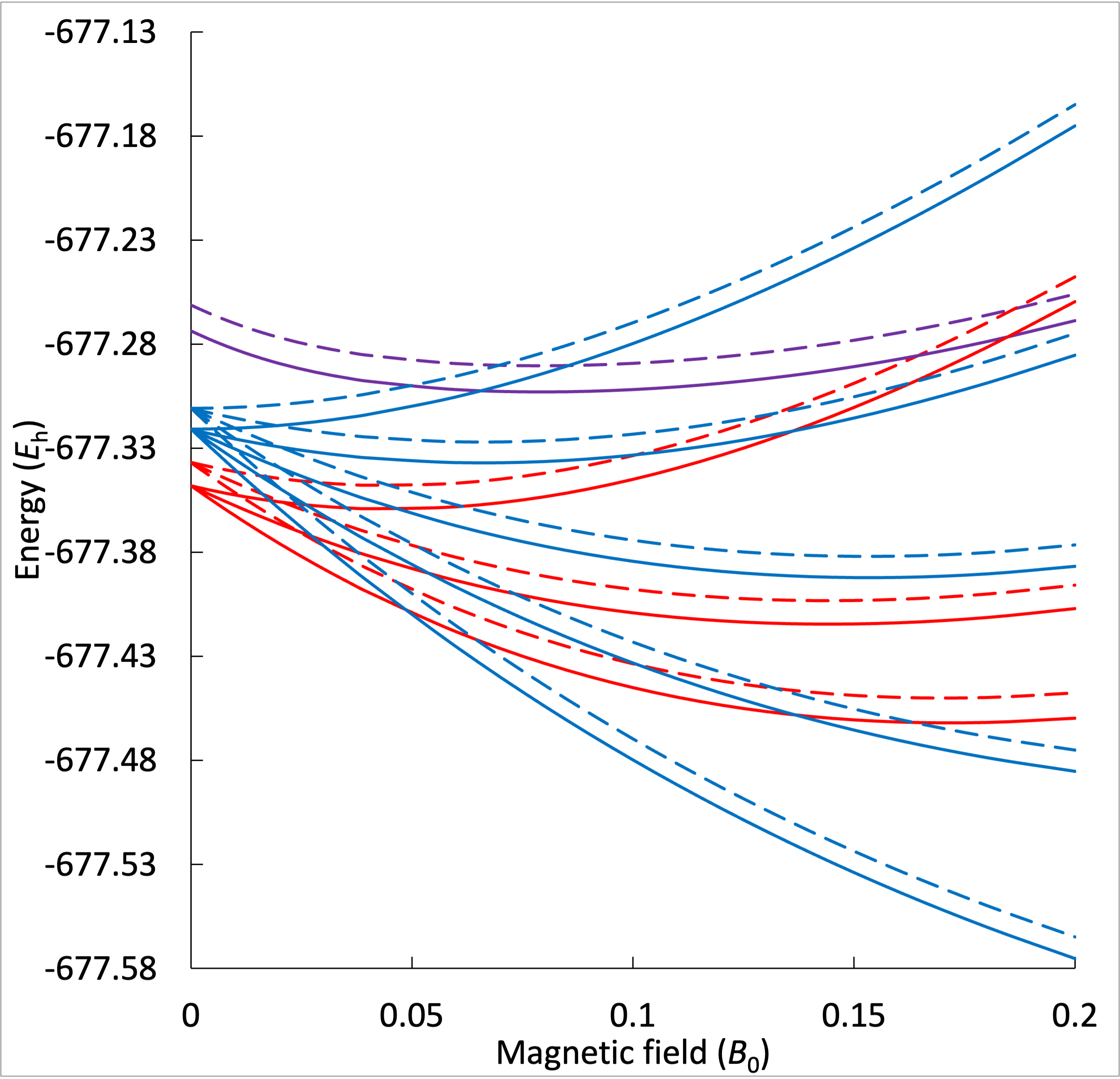}};
    \begin{scope}[x={(image.south east)},y={(image.north west)}]
        
        \node[anchor=west,minimum height = 0.5 cm] at (0.25,0.2)  {\scalebox{0.8}{CCSD}};
        \draw[dashed] (0.25,0.2) -- +(-0.06,0) node [midway] (L) {};
        \node[anchor=west,minimum height = 0.5 cm] at (0.25,0.15)  {\scalebox{0.8}{CCSD(T)(a)*}};
        \draw[-] (0.25,0.15) -- +(-0.06,0) node [midway] (L) {};

        \node[anchor=west,minimum height = 0.5 cm] at (0.25,0.9)  {\scalebox{0.8}{$^3S_g$}};
        \draw[-,color=violet] (0.25,0.9) -- +(-0.06,0) node [midway] (L) {};
        \node[anchor=west,minimum height = 0.5 cm] at (0.25,0.85)  {\scalebox{0.8}{$^3D_g$}};
        \draw[-,color=NavyBlue] (0.25,0.85) -- +(-0.06,0) node [midway] (L) {};
        \node[anchor=west,minimum height = 0.5 cm] at (0.25,0.8)  {\scalebox{0.8}{$^3P_u$}};
        \draw[-,color=red] (0.25,0.8) -- +(-0.06,0) node [midway] (L) {};
    \end{scope}
\end{tikzpicture}
\caption{\footnotesize Low-lying triplet states of Ca calculated  at the SF-EOM-CCSD and  SF-EOM-CCSD(T)(a)*  levels with unc-aug-cc-pCV5Z basis set. }
\label{fig:Ca_trip}
\end{minipage}

\begin{minipage}[t]{0.48\textwidth}
    \begin{tikzpicture}
    \node[anchor=south west,inner sep=0] (image) at (0,0) {\includegraphics[width=\textwidth]{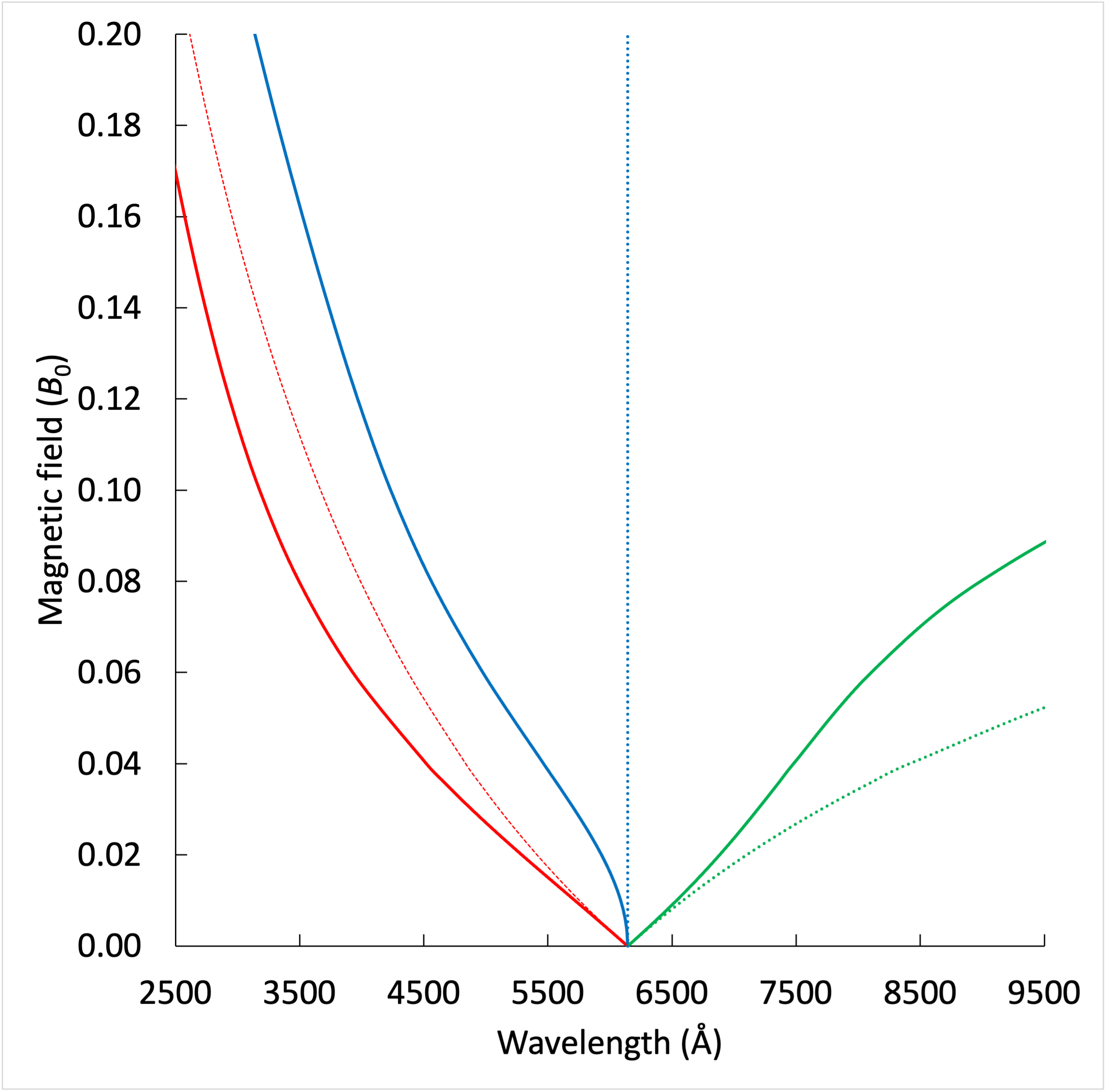}};
    \begin{scope}[x={(image.south east)},y={(image.north west)}]
        
        \node[anchor=west,minimum height = 0.5 cm] at (0.7,0.9)  {\scalebox{0.8}{Orbital-Zeeman}};
        \draw[dotted] (0.7,0.9) -- +(-0.06,0) node [midway] (L) {};
        \node[anchor=west,minimum height = 0.5 cm] at (0.7,0.85)  {\scalebox{0.8}{Extrapolation}};
        \draw[-] (0.7,0.85) -- +(-0.06,0) node [midway] (L) {};

        \node[anchor=west,minimum height = 0.5 cm] at (0.7,0.75)  {\scalebox{0.8}{$\Delta M_L=+1$}};
        \draw[-,color=BrickRed] (0.7,0.75) -- +(-0.06,0) node [midway] (L) {};
        \node[anchor=west,minimum height = 0.5 cm] at (0.7,0.7)  {\scalebox{0.8}{$\Delta M_L=0$}};
        \draw[-,color=NavyBlue] (0.7,0.7) -- +(-0.06,0) node [midway] (L) {};
        \node[anchor=west,minimum height = 0.5 cm] at (0.7,0.65)  {\scalebox{0.8}{$\Delta M_L=-1$}};
        \draw[-,color=Green] (0.7,0.65) -- +(-0.06,0) node [midway] (L) {};
    \end{scope}
\end{tikzpicture}
\caption{\footnotesize The extrapolated B-$\lambda$ curves for the $^3P_u  \rightarrow {^3S}_g$ transitions of Ca. The dotted lines correspond to results that assume a simple orbital Zeeman dependence of the energy instead of finite-field predicitions.  }
\label{fig:Ca_bl}
\end{minipage}
\hfill
\begin{minipage}[t]{0.48\textwidth}
    \begin{tikzpicture}
    \node[anchor=south west,inner sep=0] (image) at (0,0) {\includegraphics[width=\textwidth]{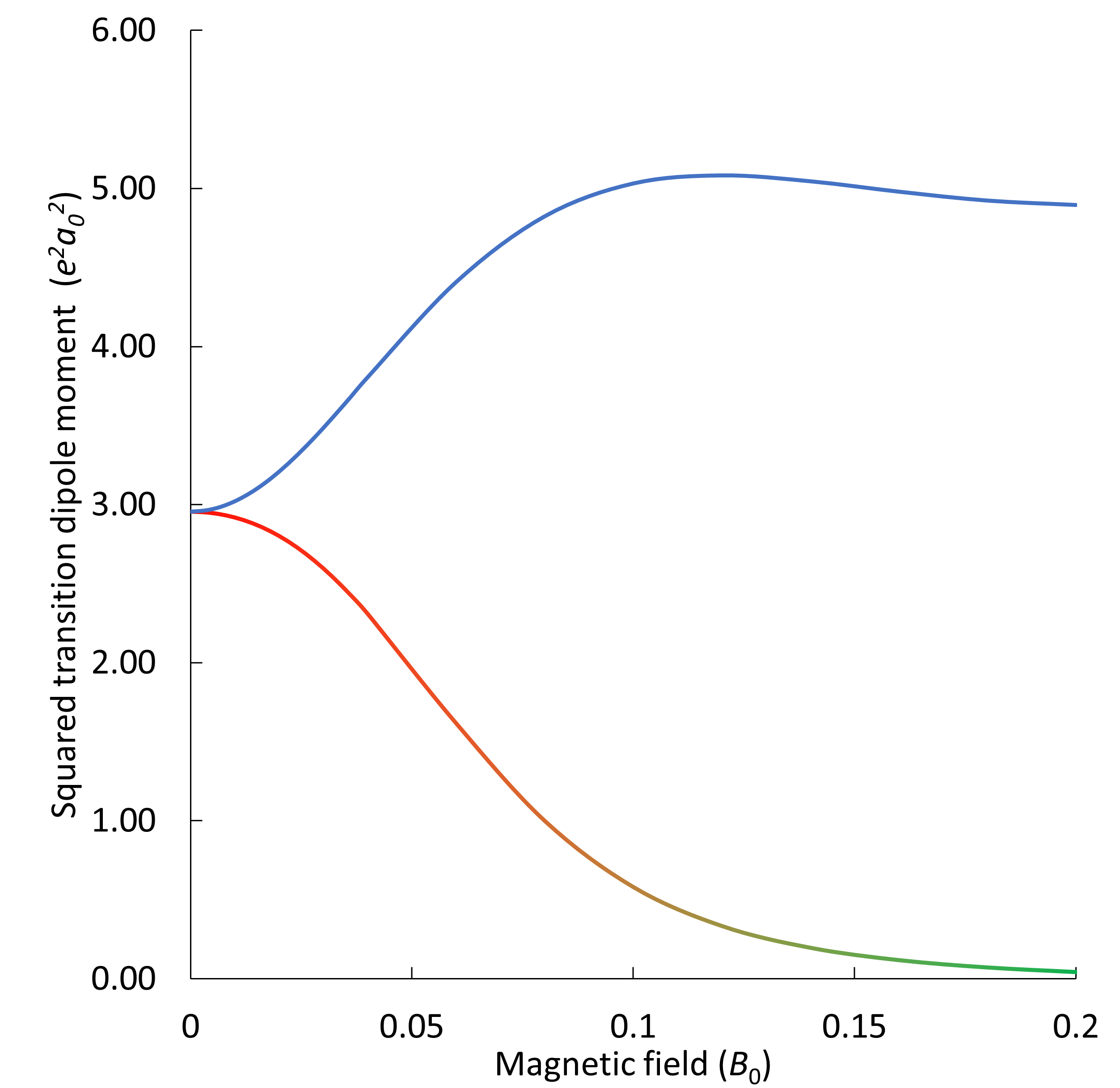}};
    \begin{scope}[x={(image.south east)},y={(image.north west)}]
        
       \node[anchor=west,minimum height = 0.5 cm] at (0.7,0.75)  {\scalebox{0.8}{$\Delta M_L=+1$}};
        \draw[-,color=BrickRed] (0.7,0.75) -- +(-0.06,0) node [midway] (L) {};
        \node[anchor=west,minimum height = 0.5 cm] at (0.7,0.7)  {\scalebox{0.8}{$\Delta M_L=0$}};
        \draw[-,color=NavyBlue] (0.7,0.7) -- +(-0.06,0) node [midway] (L) {};
        \node[anchor=west,minimum height = 0.5 cm] at (0.7,0.65)  {\scalebox{0.8}{$\Delta M_L=-1$}};
        \draw[-,color=Green] (0.7,0.65) -- +(-0.06,0) node [midway] (L) {};
    \end{scope}
\end{tikzpicture}
\caption{\footnotesize The squared transition dipole moments for the $^3P_u  \rightarrow {^3S}_g$ transitions of Ca at the SF-EOM-CCSD  level with unc-aug-cc-pCV5Z basis set. The $\Delta M_L =\pm 1$ transition have exactly the same transition dipole moment. }
\label{fig:Ca_trans}
\end{minipage}

\end{figure}

In this section, the low-lying  $^3P_u$ ([Ar]4s4p), $^3D_g$ ([Ar]4s3d), and  $^3S_g$ ([Ar]4s5s) states are studied for magnetic field strengths up to $B=0.2\BB$. 
Starting from the closed-shell $^1S_g$ ([Ar]3s$^2$) state as the CC reference, the triplet states were targeted using the SF-EOM-CC approach. The sph-unc-aug-cc-pCV$X$Z basis sets, with $X=\tn{T, Q, 5}$ were used and approximate triples corrections were accounted for at the SF-EOM-CCSD(T)(a)* level of theory.

In \figref{fig:Ca_trip}, the electronic energies of the triplet states of Ca are plotted as a function of the magnetic field. In contrast to the lighter Mg atom however, the energetically second excited state of triplet multiplicity is the $^3D_g$ state, which arises from excitations to the empty inner $3d$ orbitals. The $^3D_g / ^3\Delta_g^-$  component of the latter becomes the ground state of the system for $B\geq0.06\BB$. 
Approximate triples corrections at the SF-EOM-CCSD(T)(a)* level of theory amount to about $\sim 10\mEh$ for the electronic energies. 
Their contribution leads to an almost parallel shift to the SF-EOM-CCSD energies in the magnetic-field strengths studied here. The contributions of the triples corrections to excitation energies are one order of magnitude smaller than for the corresponding total energies, i.e., about $\sim 1\mEh$. 
One may expect that the states are accurately described at this level of theory as no predominant double-excitation character is observed neither for the field strengths studied here nor in field-free studies on the performance of the CCSD(T)(a)* approach.\cite{Matthews2020a,Verplancke2023}

 For the Ca atom, the electronic transitions of interest for WD spectra are  between the $^3P_u$ and $^3S_g$ states. The corresponding squared transition dipole moments (STMs) $|\mu_{I\rightarrow J}|^2= \mu_{I\rightarrow J}\mu_{J \rightarrow I}$ are shown as a function of the magnetic field in Fig.~\ref{fig:Ca_trans}. While for stronger fields of $0.2\BB$ the transitions from the $ p_{\pm{1}}$ orbitals go to zero, the STM for a transition from the $p_0$ orbital is increased and will lead to more intense signals. As already discussed in Ref.~\onlinecite{Hampe2019} for the $s\rightarrow p$ transitions of sodium, this behavior can be explained by the fact that the orbitals oriented perpendicular to the magnetic field ($p_{\pm{ 1}}$) are contracted when the field is increased while the $p_0$ orbital, oriented parallel to the field is stretched along the field direction leading to a larger overlap with the $s$ orbital. The effect is more pronounced here since the involved $s$ orbital is more diffuse and hence a larger overlap is achieved. A similar situation occurs for Mg ($^3P_u \rightarrow ^3S_g$) transitions, see also Ref.~\onlinecite{Blaschke2024}, Fig. S5 in the SI. 
 While the $^3P_u\rightarrow^3D_g$ transitions exhibit large oscillator strengths in our calculations, they are outside the relevant wavelength window typically measured in WD spectra and they are also not reported in the NIST database.\cite{NIST_ASD} As such, they are not considered further.
 Following the extrapolation scheme described in Ref.~\onlinecite{Hampe2020}, B-$\lambda$ curves have been generated which are presented in \figref{fig:Ca_bl} for the $^3P_u\rightarrow^3S_g$ transition. The zero-field shift correction relative to reference data from the NIST database\cite{NIST_ASD} amounts to $10\AA$. 
For the middle component, a similar trend as for the Mg atom is observed. Because of the deformation of the $5s$ orbital and its mixing with d-type orbitals, the $\Delta M_L=0$ excitation strongly deviates from the orbital-Zeeman splitting.\cite{Hollands2023,Kitsaras2024,Kitsarasthes}

It is important to point out that the use of the SF-EOM-CC approach in combination with a closed-shell CC reference results in open-shell triplet state wavefunctions free of spin-contamination.  Moreover, the EOM-CCSD(T)(a)* approach allows for the calculation of triples corrections in systems for which a full EOM-CCSDT treatment is not feasible. In addition, due to the non-iterative nature, it is more efficient compared to the iterative $M^7$ EOM-CC3 approach which has been used in previous studies.\cite{Kitsaras2024}\footnote{Note that for the prediction of intensities, transition-dipole moments have been calculated for Na, Mg, and Ca in Ref.~\onlinecite{Hollands2023}. Furthermore, in the SI in section III,  transition-dipole moments for the $2s \rightarrow 2p$ transition of sodium were calculated using the EA-EOM-CC approach and compared against EE-EOM-CCSD results using from Ref.~\onlinecite{Hampe2019}. The results are essentially identical. }

%%%%%%%%%%%%%%%%%%%%%%%%%%%%%%%%%%%%%%%%%%%%%%%%%%%%%%%%%%%%%%%%%%%%%%%%%%%%
\subsection{Electronic excitations and properties of the CH radical}
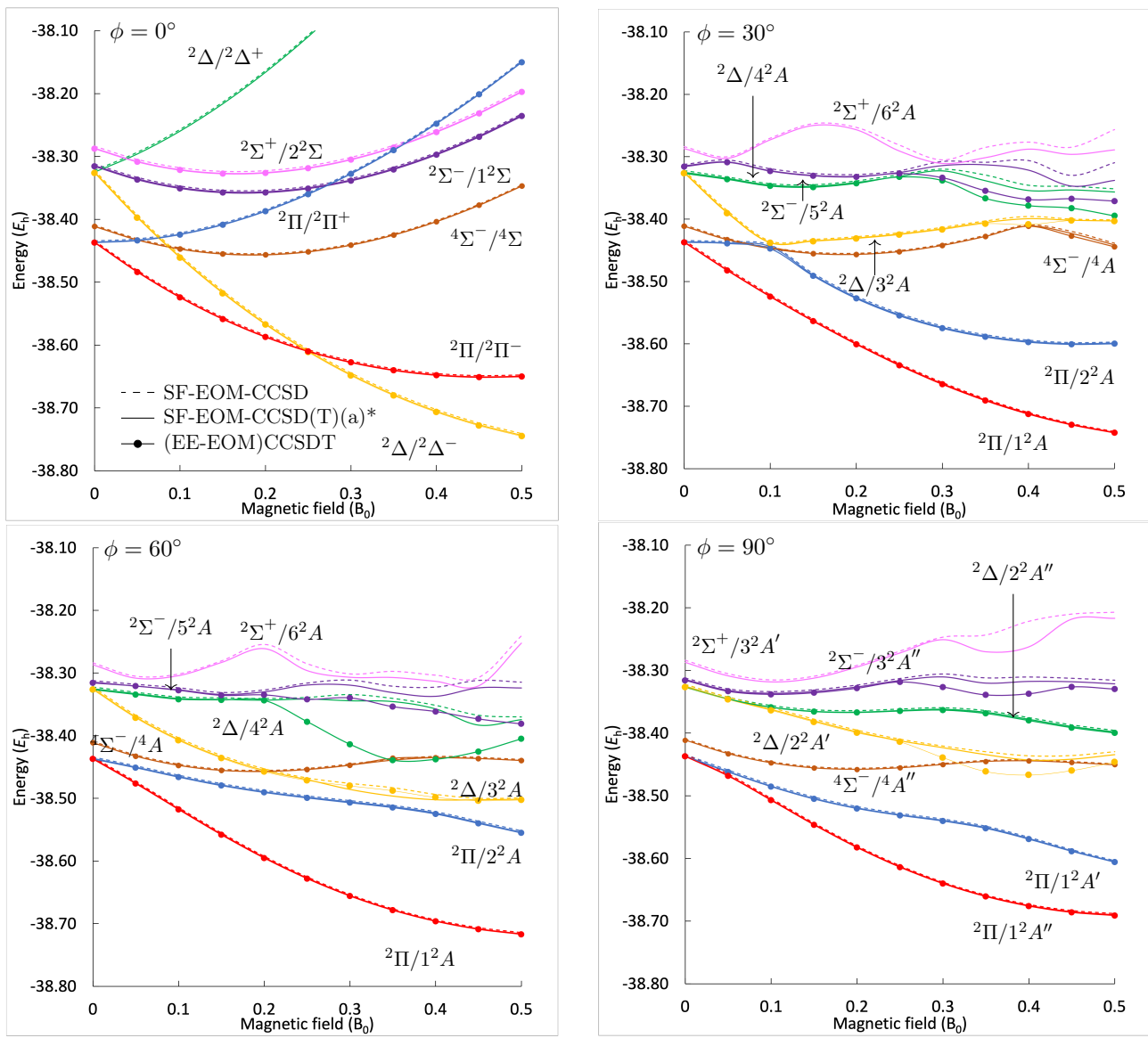
\begin{figure}[b!]
\centering
\begin{minipage}{\textwidth}
    \begin{minipage}[t]{0.48\textwidth}
    \begin{tikzpicture}
        \node[anchor=south west,inner sep=0] (image) at (0,0) {\includegraphics[width=\textwidth]{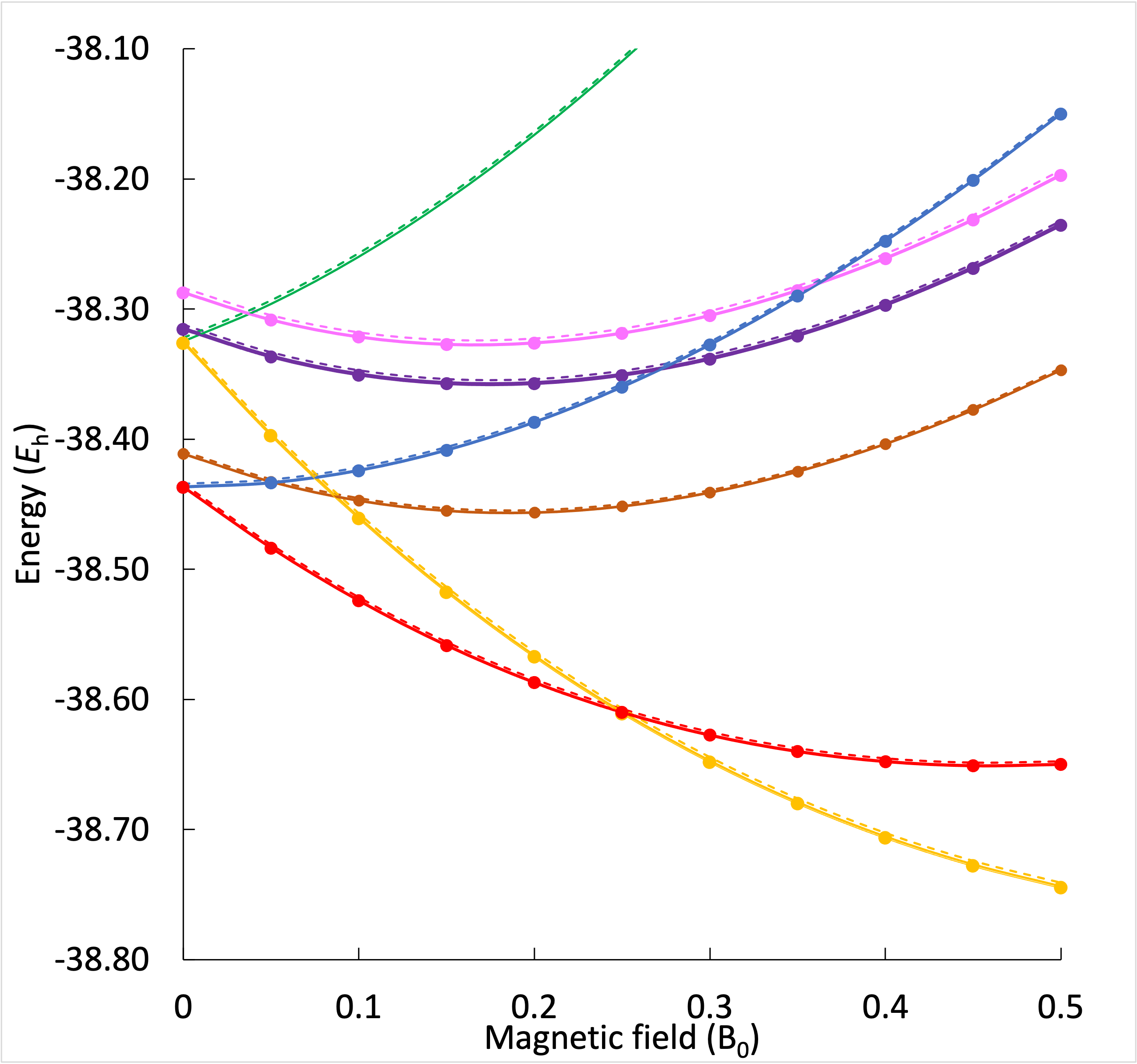}};
        \begin{scope}[x={(image.south east)},y={(image.north west)}]
            \node at (0.25,0.95) (angle) {\scalebox{0.9}{$\phi=0^\circ$}};
            \node at (0.87,0.33) (label1) {\scalebox{0.8}{$^2\Pi/^2\Pi^-$}};
            \node at (0.56,0.58) (label2) {\scalebox{0.8}{$^2\Pi/^2\Pi^+$}};
            \node at (0.87,0.55) (label3) {\scalebox{0.8}{${^4\Sigma}^- /{^4\Sigma}$}};
            \node at (0.75,0.14) (label4) {\scalebox{0.8}{$^2\Delta/^2\Delta^-$}};
            \node at (0.4,0.9) (label5) {\scalebox{0.8}{$^2\Delta/^2\Delta^+$}};
            \node at (0.84,0.67) (label6) {\scalebox{0.8}{$^2\Sigma^-/1{^2\Sigma}$}};
            \node at (0.5,0.72) (label7) {\scalebox{0.8}{${^2\Sigma}^+/2{^2\Sigma}$}};

            \node[anchor=west,minimum height = 0.5 cm] at (0.27,0.25)  {\scalebox{0.8}{SF-EOM-CCSD}};
            \draw[dashed] (0.27,0.25) -- +(-0.06,0) node [midway] (L) {};
            \node[anchor=west,minimum height = 0.5 cm] at (0.27,0.2)  {\scalebox{0.8}{SF-EOM-CCSD(T)(a)*}};
            \draw[-] (0.27,0.2) -- +(-0.06,0) node [midway] (L) {};
            \node[anchor=west,minimum height = 0.5 cm] at (0.27,0.15)  {\scalebox{0.8}{(EE-EOM)CCSDT}};
            \draw[-] (0.27,0.15) -- +(-0.06,0) node [midway] (L) {};
            \filldraw[black] (L)  circle  (1.5pt);

        \end{scope}
    \end{tikzpicture}
    \end{minipage} \hspace{0.02\textwidth}
    \begin{minipage}[t]{0.48\textwidth}
    \begin{tikzpicture}
        \node[anchor=south west,inner sep=0] (image) at (0,0) {\includegraphics[width=\textwidth]{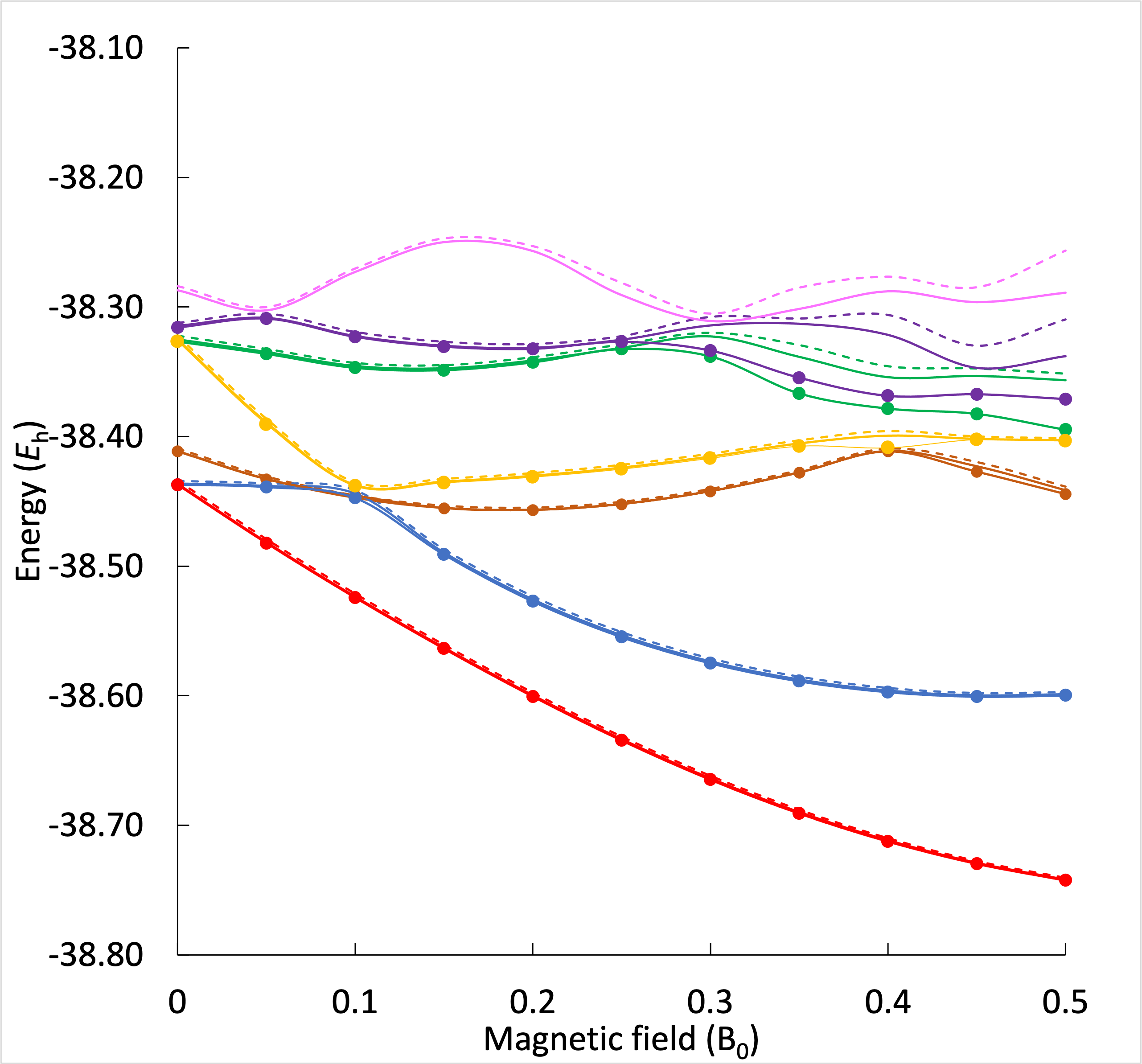}};
        \begin{scope}[x={(image.south east)},y={(image.north west)}]
            \node at (0.25,0.95) (angle) {\scalebox{0.9}{$\phi=30^\circ$}};
            \node at (0.75,0.15) (label1) {\scalebox{0.8}{$^2\Pi/1^2A $}};
            \node at (0.87,0.28) (label2) {\scalebox{0.8}{$^2\Pi/2^2A $}};
            \node at (0.87,0.5) (label3) {\scalebox{0.8}{$ {^4\Sigma}^-/{^4A}$}};
            \node at (0.5,0.46) (label4) {\scalebox{0.8}{$^2\Delta/3^2A$}};
            \draw[->] (label4)+(-0.0,0.02) -- +(-0.0,0.09);
            \node at (0.28,0.87) (label5) {\scalebox{0.8}{$^2\Delta/4^2A $}};
            \draw[->] (label5) -- +(-0.0,-0.2);
            \node at (0.37,0.605) (label6) {\scalebox{0.8}{$ {^2\Sigma}^-/5{^2A}$}};
            \draw[->] (label6)+(-0.0,0.02) -- +(-0.0,0.062);
            \node at (0.5,0.8) (label7) {\scalebox{0.8}{${^2\Sigma}^+/6{^2A}$}};

        \end{scope}
    \end{tikzpicture}
    \end{minipage}
\end{minipage}

\begin{minipage}{\textwidth}
    \begin{minipage}[t]{0.48\textwidth}
    \begin{tikzpicture}
        \node[anchor=south west,inner sep=0] (image) at (0,0) {\includegraphics[width=\textwidth]{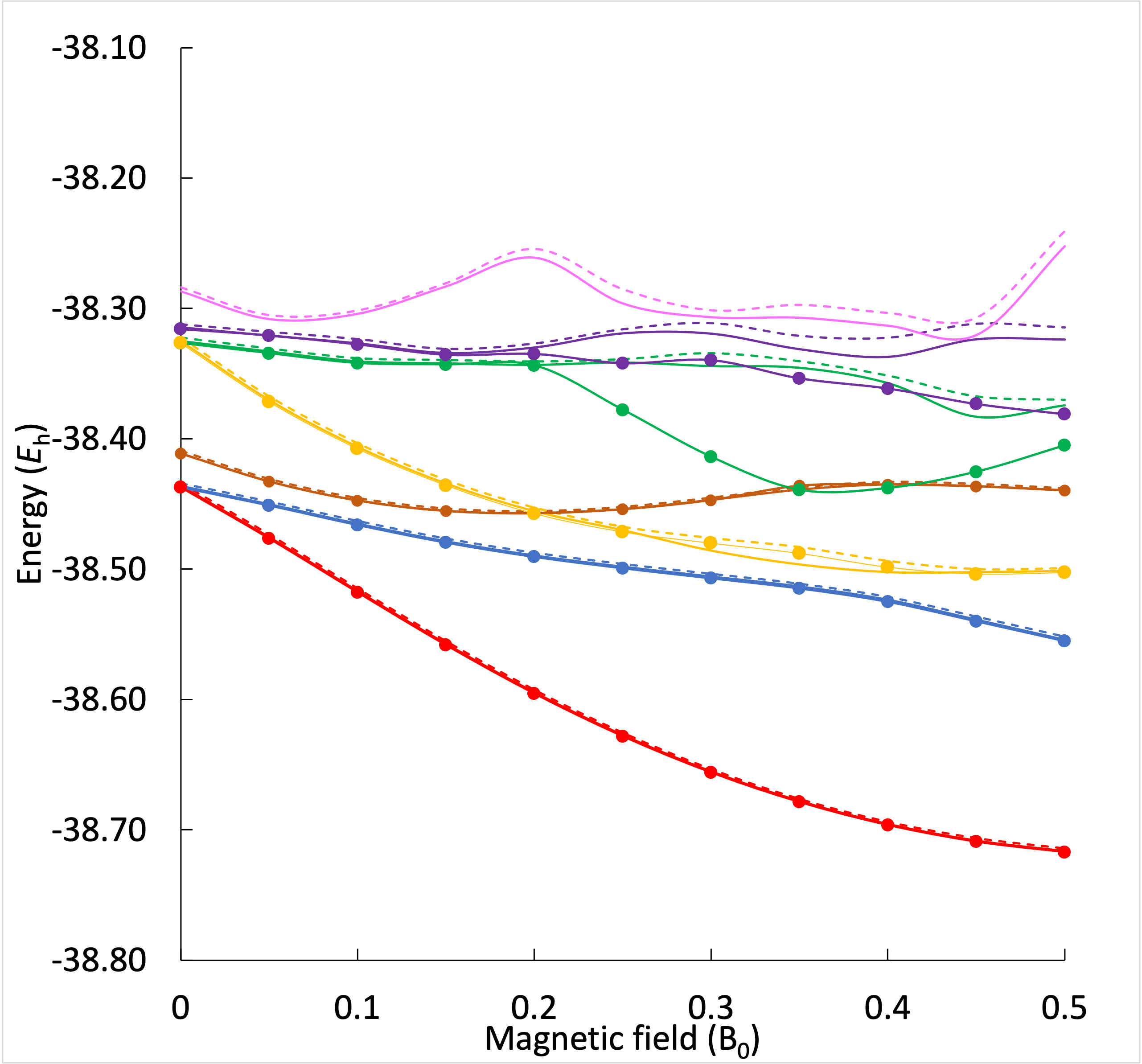}};
        \begin{scope}[x={(image.south east)},y={(image.north west)}]
            \node at (0.25,0.95) (angle) {\scalebox{0.9}{$\phi=60^\circ$}};
            \node at (0.75,0.15) (label1) {\scalebox{0.8}{$^2\Pi/1^2A $}};
            \node at (0.87,0.35) (label2) {\scalebox{0.8}{$^2\Pi/2^2A $}};
            \node at (0.22,0.57) (label3) {\scalebox{0.8}{$ {^4\Sigma}^-/{^4A}$}};
            \node at (0.87,0.48) (label4) {\scalebox{0.8}{$^2\Delta/3^2A$}};
            \node at (0.44,0.6) (label5) {\scalebox{0.8}{$^2\Delta/4^2A $}};
            \node at (0.3,0.8) (label6) {\scalebox{0.8}{$ {^2\Sigma}^-/5{^2A}$}};
            \draw[->] (label6) -- +(0.0,-0.124);
            \node at (0.5,0.79) (label7) {\scalebox{0.8}{${^2\Sigma}^+/6{^2A}$}};
        \end{scope}
    \end{tikzpicture}
    \end{minipage} \hspace{0.02\textwidth}
    \begin{minipage}[t]{0.48\textwidth}
    \begin{tikzpicture}
        \node[anchor=south west,inner sep=0] (image) at (0,0) {\includegraphics[width=\textwidth]{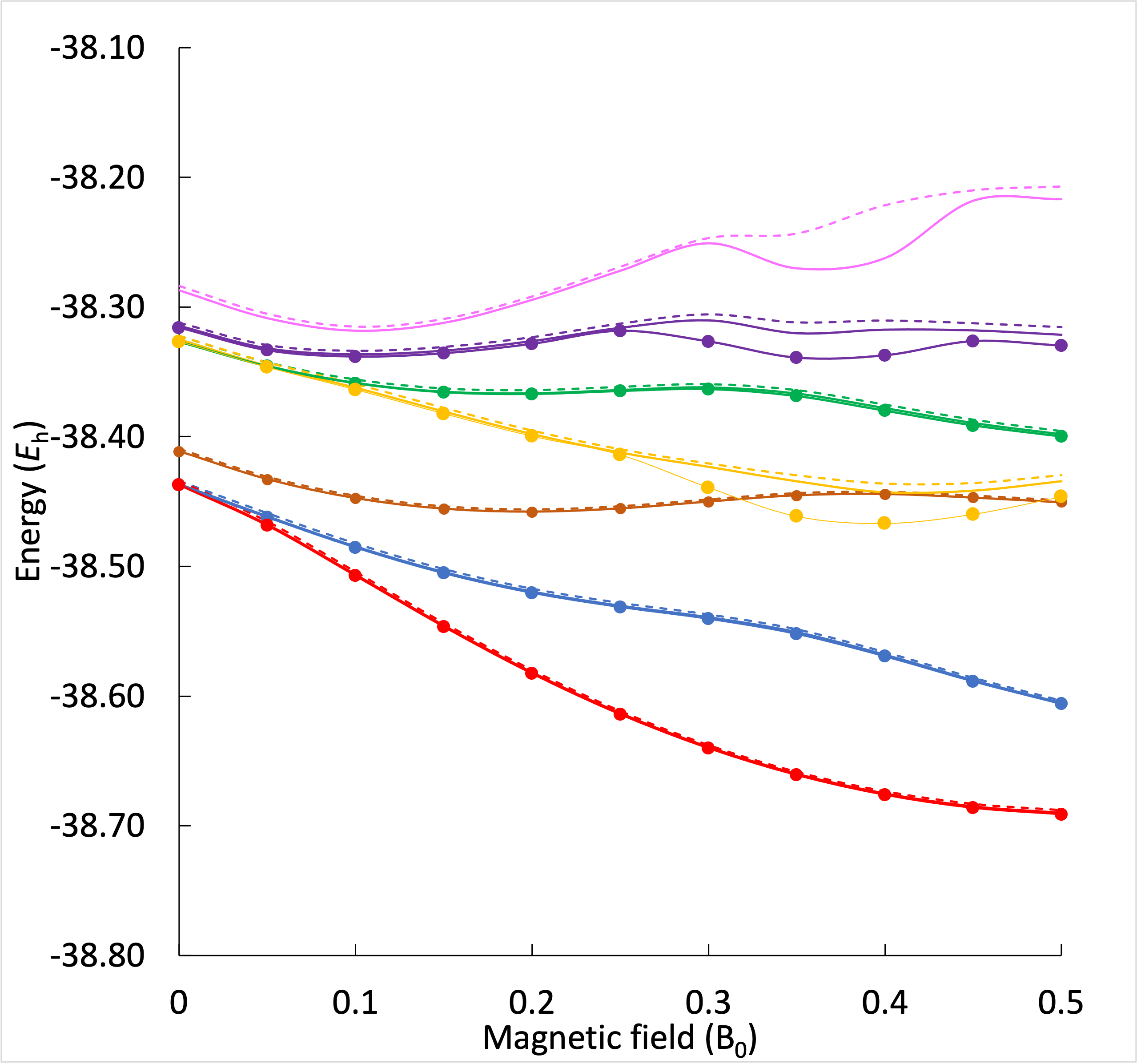}};
        \begin{scope}[x={(image.south east)},y={(image.north west)}]
            \node at (0.25,0.95) (angle) {\scalebox{0.9}{$\phi=90^\circ$}};
            \node at (0.75,0.2) (label1) {\scalebox{0.8}{$^2\Pi/1^2A''$}};
            \node at (0.84,0.3) (label2) {\scalebox{0.8}{$^2\Pi/1^2A'$}};
            %\draw[->] (label2) -- +(-0.0,0.25);
            \node at (0.5,0.49) (label3) {\scalebox{0.8}{${^4\Sigma}^-/{^4A}'' $}};
            %\draw[->] (label2) -- +(-0.0,0.25);
            \node at (0.35,0.57) (label4) {\scalebox{0.8}{$^2\Delta/2^2A' $}};
            %\draw[->] (label4)+(-0.0,0.02) -- +(-0.0,0.07);
            \node at (0.75,0.9) (label5) {\scalebox{0.8}{$^2\Delta/2^2A'' $}};
            \draw[->] (label5) -- +(-0.0,-0.28);
            \node at (0.5,0.73) (label6) {\scalebox{0.8}{${^2\Sigma}^-/3{^2A}'' $}};
            %\draw[->] (label6)+(-0.0,0.02) -- +(0.01,0.055);
            \node at (0.25,0.76) (label7) {\scalebox{0.8}{$ {^2\Sigma}^+/3{^2A}'$}};

        \end{scope}
    \end{tikzpicture}
    \end{minipage}
\end{minipage}
\caption{The low-lying doublet states and a quartet state ($M_S=-\frac{1}{2}$) of CH in different magnetic-field orientations at the  SF-EOM-CCSD  and SF-EOM-CCSD(T)(a)* levels of theory compared to EE-EOM-CCSDT results obtained using the unc-aug-cc-pCVDZ basis set. }
\label{fig:CH_SF}
\end{figure}

Due to the fact that CH has been detected in weakly magnetic WDs,\cite{Berdyugina2007,Vornanen2010,Vornanen2013} it is reasonable to assume that it also occurs on strongly magnetized WDs. The evolution of its electronic spectrum as a function of the magnetic field has been studied using the standard EE-EOM-CC flavor at the CC2, CCSD, CC3, and CCSDT levels of theory.\cite{Kitsaras2024}  It turned out that despite the simplicity of the molecule, the evolution of the electronic structure of the excited states is rather demanding. Symmetry-breaking due to unequal handling of degenerate states in the field-free limit, spin-contamination and double-excitation character were among the challenges for  treating the system in a consistent manner.\cite{Kitsaras2024} 
More concretely, using EE-EOM-CCSD, only the $^2\Pi$ states could be consistently described in a satisfactory manner with a deviation of $1 \mEh$ relative to the CCSDT results.
The other states of interest, i.e., the $^2\Delta$, the $^2\Sigma^-$, and $^2\Sigma^+$ states (cf. Fig.~\ref{fig:CH_SF}) are either not easily targeted, difficult to characterize, or strongly spin contaminated and moreover are accompanied with a large deviation relative to CCSDT reaching $10 \mEh$ due to strong double-excitation character.\cite{Kitsarasthes}
The $^2\Sigma^+$ state was for example not found in the EE-EOM-CC calculations at all for non-parallel orientations of the magnetic field probably due to strong mixing with the $^2\Delta$ state and other higher-lying configurations.
A more detailed discussion can be found in Ref.~\onlinecite{Kitsarasthes}.

Here, we attempt to remedy some of the difficulties mentioned above by using the flexible arsenal of  different EOM-CC variants and discuss their limitations. Specifically, we use the following two approaches: a) We access the degenerate $^2\Pi$ ground state using an EA-EOM-CC treatment starting from the $^1\Sigma^+$ ground state of the cation as a CC reference state. This approach yields spin pure states which are also free of symmetry breaking. However, low-lying excited states of the neutral system are not well described by EA-EOM-CC as they would have significant double excitation character. We hence omit the calculation of these states with this protocol. b) We treat the low-lying $^2\Pi$, $^2\Delta$, $^2\Sigma^-$, and $^2\Sigma^+$ states at the SF-EOM-CC level starting from the $^4\Sigma^-$ state. This approach gives results with low spin-contamination as the reference is a high-spin open-shell state dominated by a single determinant with no symmetry breaking. Most importantly, using this protocol, all states of interest to us have a predominant single-excitation character at least in the field-free limit. Triples corrections are accounted for at the EOM-CCSD(T)(a)* level of theory. For the calculations, a C-H distance of $2.1410\Bohr$ was used and the sph-unc-aug-cc-pCVDZ basis set was employed. The results are compared against those obtained at the (EE-EOM-)CCSDT level of theory which serve as a reference. The magnetic field was applied at the following angles relative to the molecular bond: $0^\circ$, $30^\circ$, $60^\circ$, and $90^\circ$ for field strengths between $0$ and $0.5\BB$.

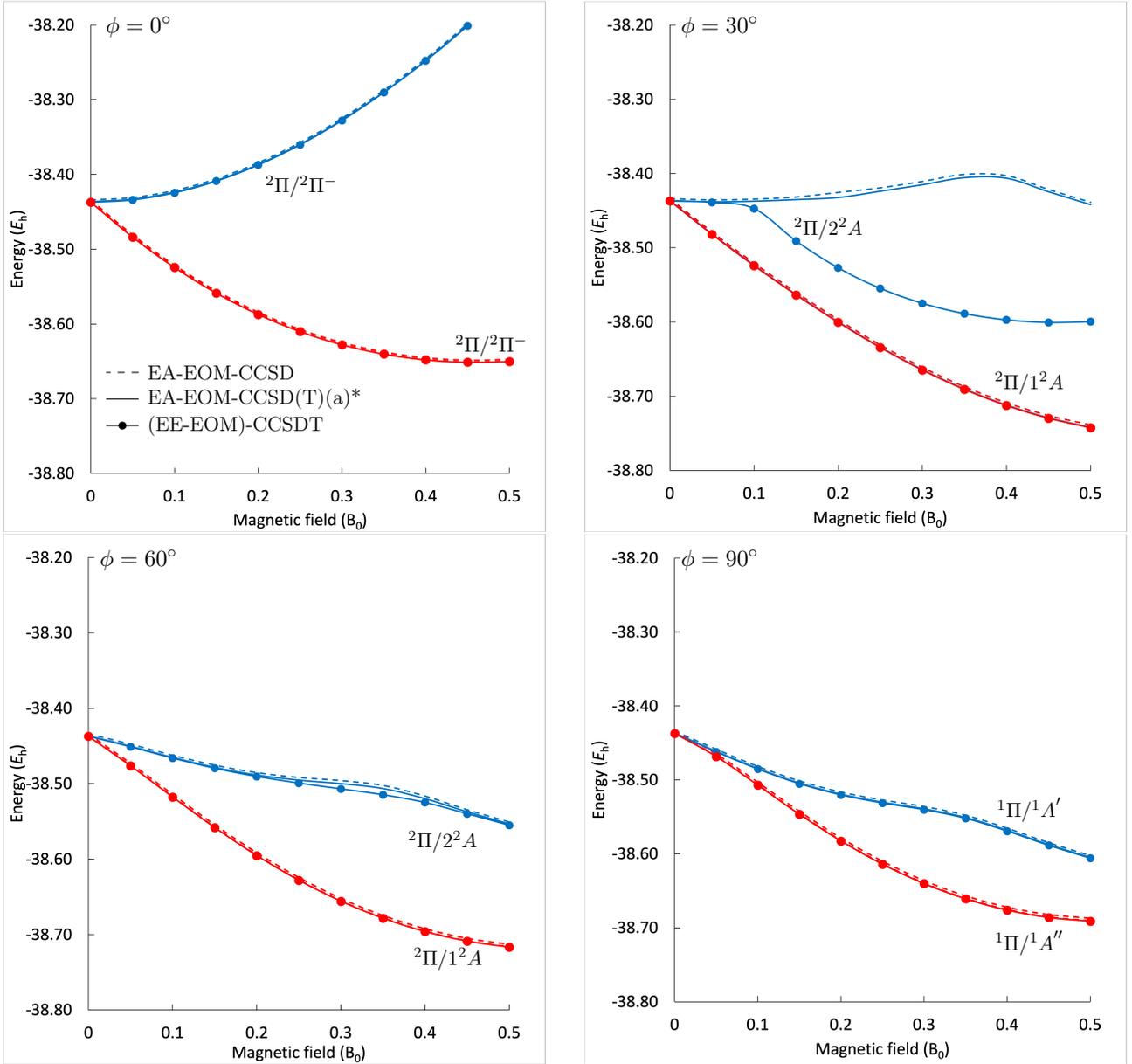
\begin{figure}[!t]
\centering
\begin{minipage}{\textwidth}
    \begin{minipage}[t]{0.48\textwidth}
    \begin{tikzpicture}
        \node[anchor=south west,inner sep=0] (image) at (0,0) {\includegraphics[width=\textwidth]{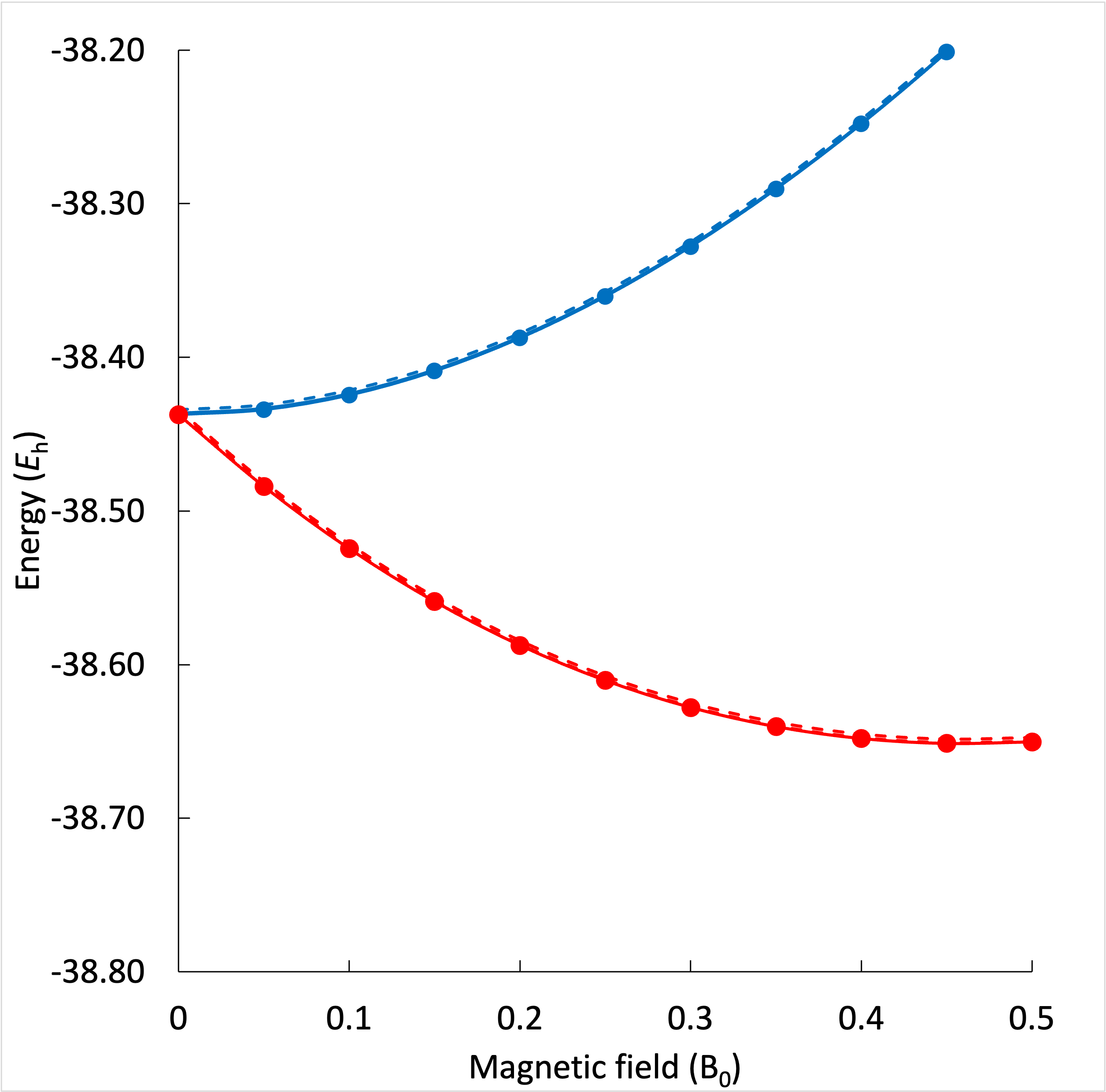}};
        \begin{scope}[x={(image.south east)},y={(image.north west)}]
            \node at (0.25,0.95) (angle) {\scalebox{0.9}{$\phi=0^\circ$}};
            \node at (0.9,0.35) (label2) {\scalebox{0.8}{$^2\Pi/^2\Pi^-$}};
            \node at (0.55,0.65) (label6) {\scalebox{0.8}{$^2\Pi/^2\Pi^-$}};

            \node[anchor=west,minimum height = 0.5 cm] at (0.25,0.3)  {\scalebox{0.8}{EA-EOM-CCSD}};
            \draw[dashed] (0.25,0.3) -- +(-0.06,0) node [midway] (L) {};
            \node[anchor=west,minimum height = 0.5 cm] at (0.25,0.25)  {\scalebox{0.8}{EA-EOM-CCSD(T)(a)*}};
            \draw[-] (0.25,0.25) -- +(-0.06,0) node [midway] (L) {};
            \node[anchor=west,minimum height = 0.5 cm] at (0.25,0.2)  {\scalebox{0.8}{(EE-EOM)-CCSDT}};
            \draw[-] (0.25,0.2) -- +(-0.06,0) node [midway] (L) {};
            \filldraw[black] (L)  circle  (1.5pt); 
        \end{scope}
    \end{tikzpicture}
    \end{minipage} \hspace{0.02\textwidth}
    \begin{minipage}[t]{0.48\textwidth}
    \begin{tikzpicture}
        \node[anchor=south west,inner sep=0] (image) at (0,0) {\includegraphics[width=\textwidth]{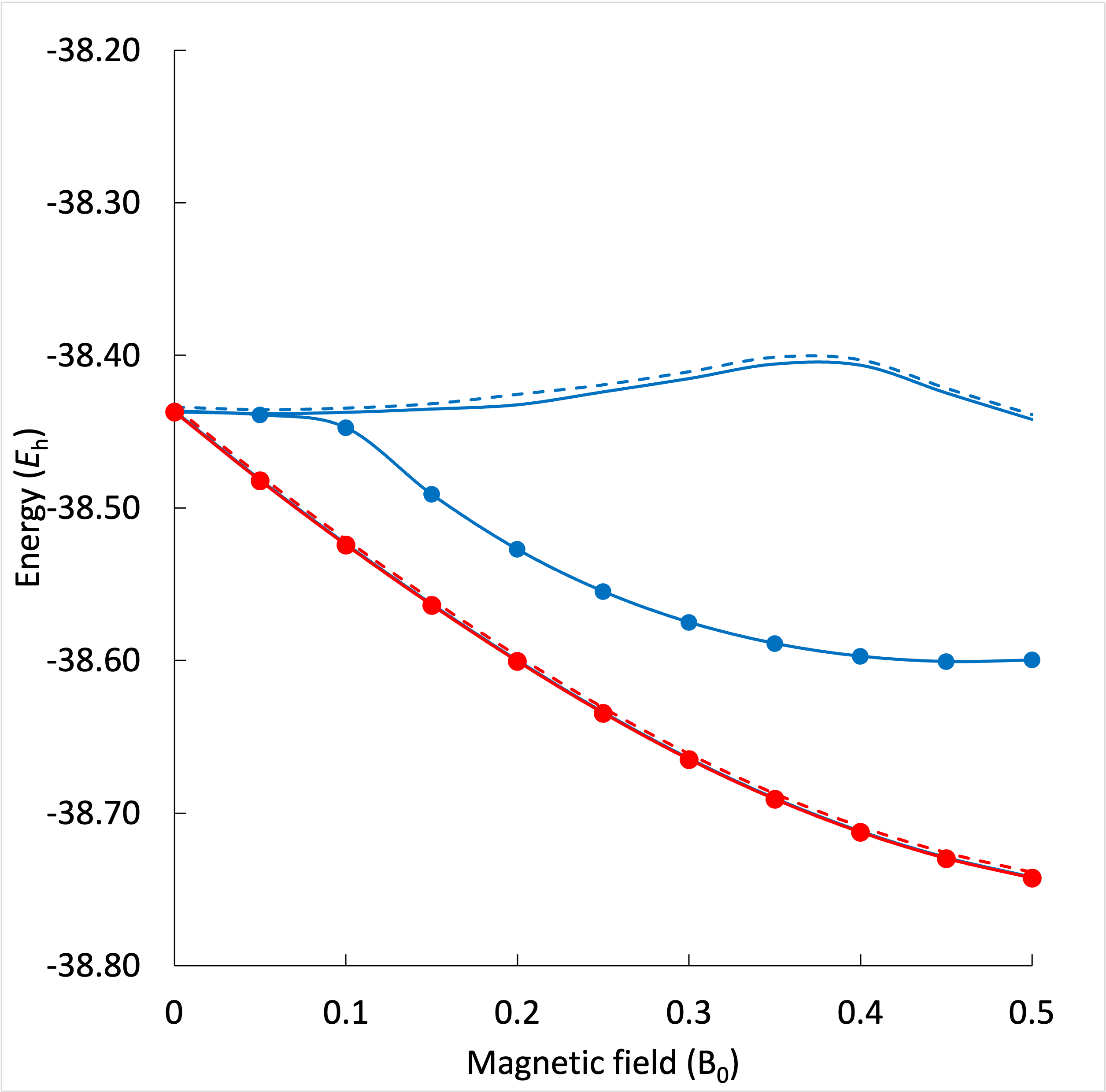}};
        \begin{scope}[x={(image.south east)},y={(image.north west)}]
            \node at (0.25,0.95) (angle) {\scalebox{0.9}{$\phi=30^\circ$}};
            \node at (0.82,0.28) (label2) {\scalebox{0.8}{$^2\Pi/1^2A $}};
            \node at (0.45,0.57) (label6) {\scalebox{0.8}{$^2\Pi/2^2A $}};
        \end{scope}
    \end{tikzpicture}
    \end{minipage}
\end{minipage}

\begin{minipage}{\textwidth}
    \begin{minipage}[t]{0.48\textwidth}
    \begin{tikzpicture}
        \node[anchor=south west,inner sep=0] (image) at (0,0) {\includegraphics[width=\textwidth]{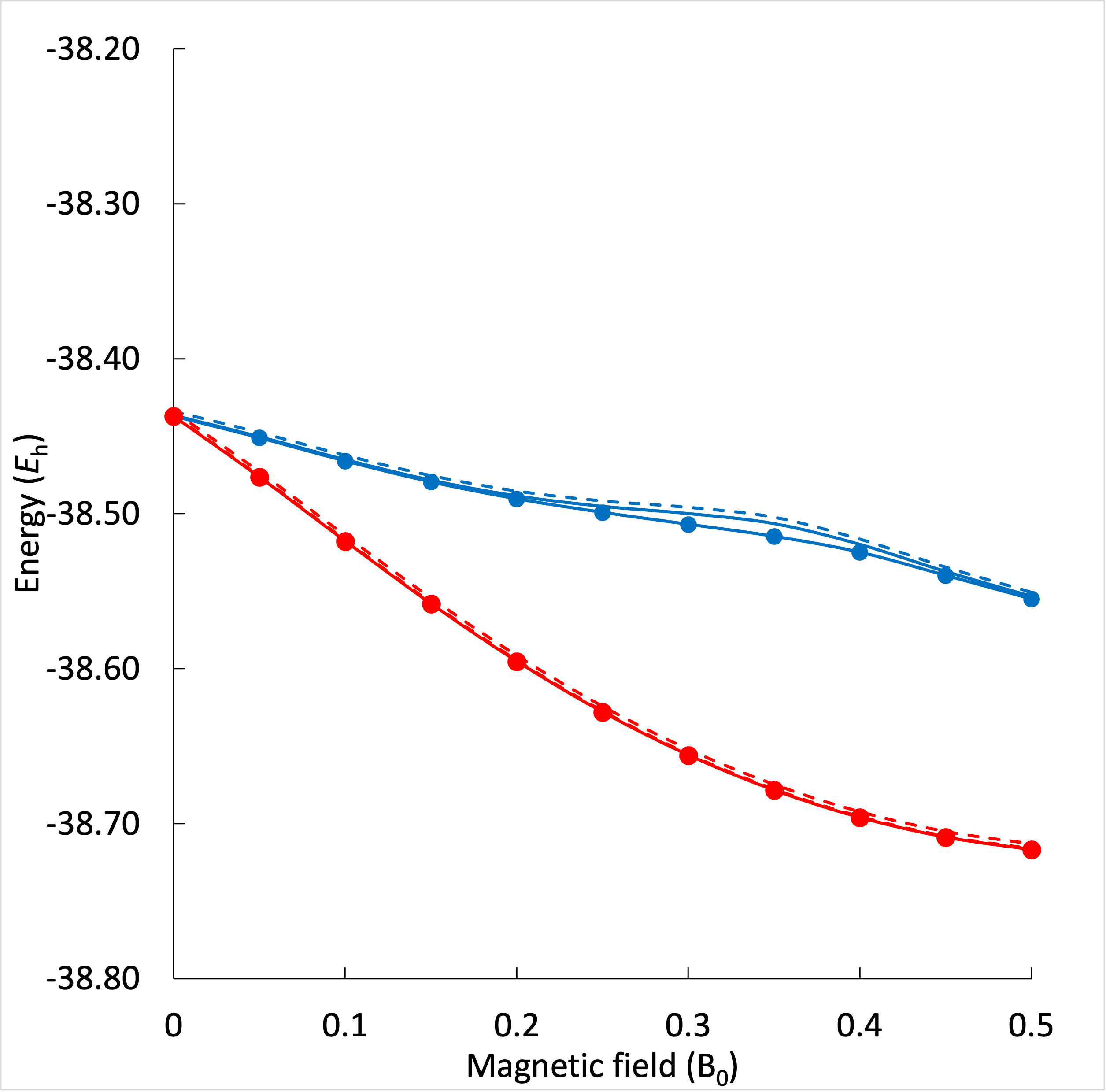}};
        \begin{scope}[x={(image.south east)},y={(image.north west)}]
            \node at (0.25,0.95) (angle) {\scalebox{0.9}{$\phi=60^\circ$}};
            \node at (0.82,0.2) (label2) {\scalebox{0.8}{$^2\Pi/1^2A $}};
            \node at (0.81,0.42) (label6) {\scalebox{0.8}{$^2\Pi/2^2A $}};
        \end{scope}
    \end{tikzpicture}
    \end{minipage} \hspace{0.02\textwidth}
    \begin{minipage}[t]{0.48\textwidth}
    \begin{tikzpicture}
        \node[anchor=south west,inner sep=0] (image) at (0,0) {\includegraphics[width=\textwidth]{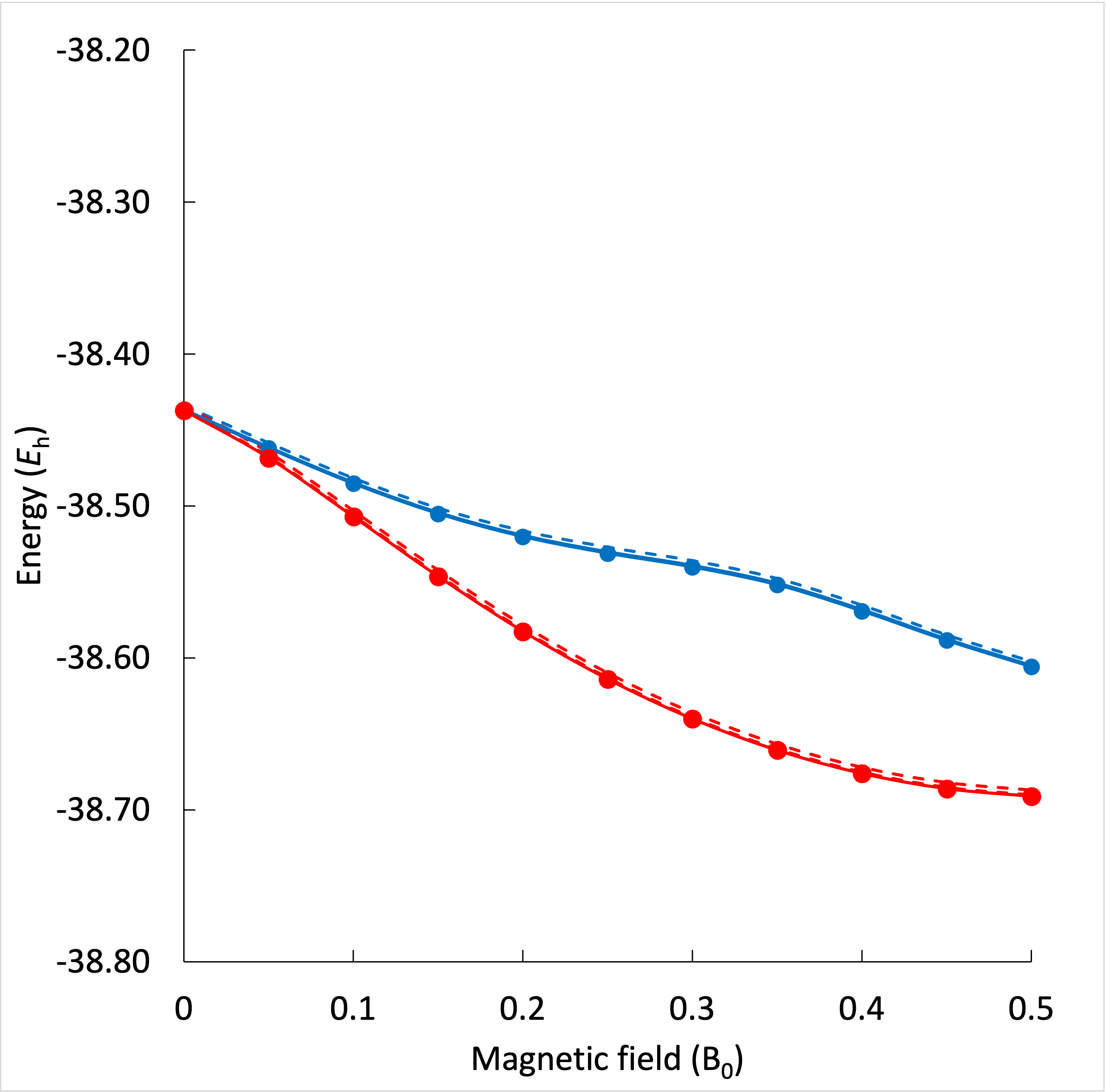}};
        \begin{scope}[x={(image.south east)},y={(image.north west)}]
            \node at (0.25,0.95) (angle) {\scalebox{0.9}{$\phi=90^\circ$}};
            \node at (0.82,0.23) (label2) {\scalebox{0.8}{$ ^1\Pi/{^1A}''$}};
            \node at (0.82,0.48) (label6) {\scalebox{0.8}{$^1\Pi/{^1A}' $}};

        \end{scope}
    \end{tikzpicture}
    \end{minipage}
\end{minipage}
\caption{ The two components of the $^2\Pi$ field-free ground state of CH in different magnetic-field orientations at the  EA-EOM-CCSD and EA-EOM-CCSD(T)(a)*  levels of theory compared to EE-EOM-CCSDT results  obtained using the unc-aug-cc-pCVDZ basis set. }
\label{fig:CH_EA}
\end{figure}

\begin{figure}

    \begin{subfigure}{0.48\textwidth}
        
        \begin{tikzpicture}
    \node[anchor=south west,inner sep=0] (image) at (0,0) {\includegraphics[width=\linewidth]{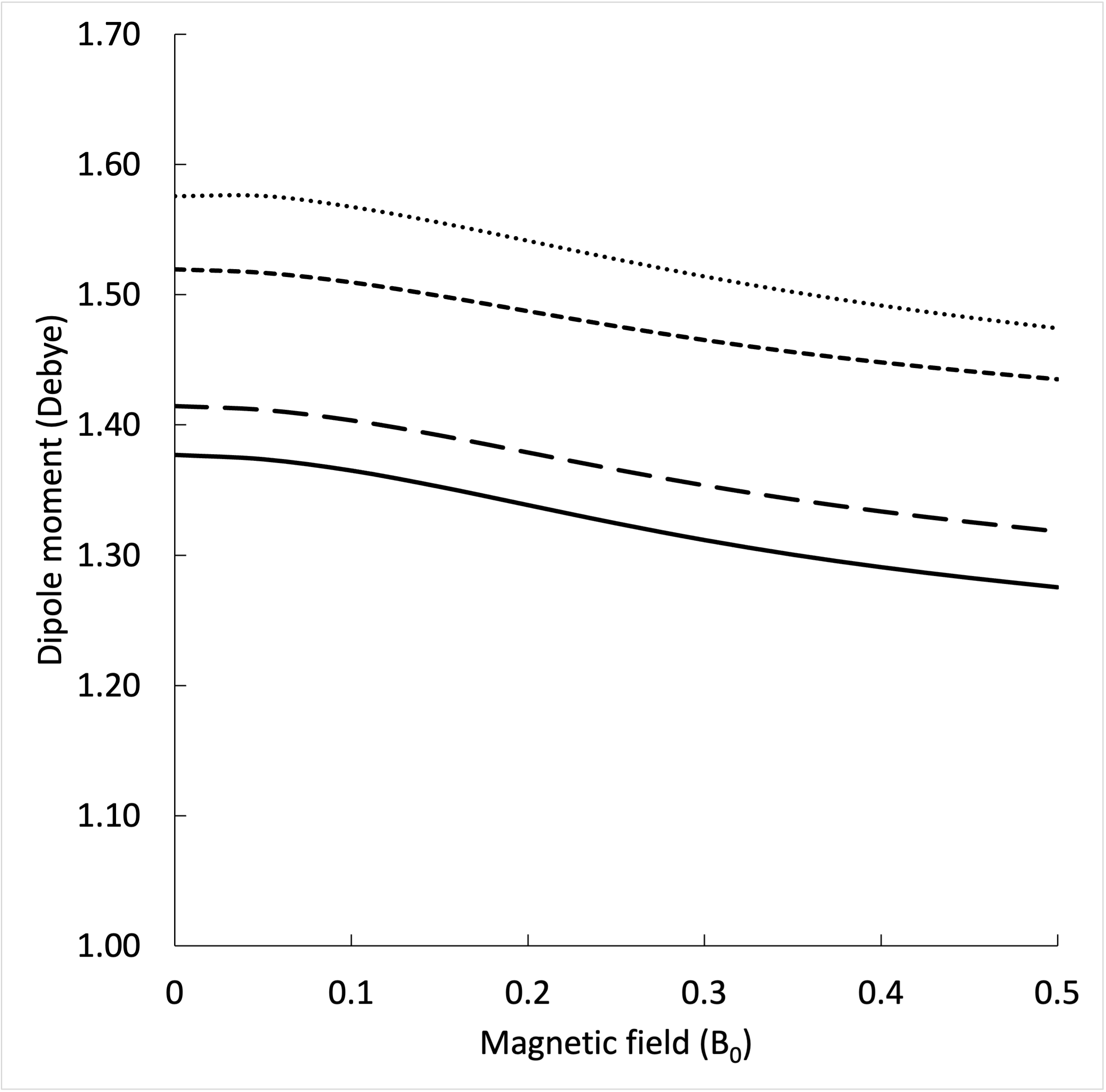}};
    \begin{scope}[x={(image.south east)},y={(image.north west)}]
        \node at (0.25,0.95) (angle) {\scalebox{0.9}{$\phi=0^\circ$}};
        
        \node[anchor=west,minimum height = 0.5 cm] at (0.25,0.35)  {\scalebox{0.8}{HF}};
        \draw[dotted] (0.25,0.35) -- +(-0.06,0) node [midway] (L) {};
        \node[anchor=west,minimum height = 0.5 cm] at (0.25,0.3)  {\scalebox{0.8}{SF-EOM-CCSD}};
        \draw[dash pattern=on 6pt off 2pt] (0.25,0.3) -- +(-0.06,0) node [midway] (L) {};
        \node[anchor=west,minimum height = 0.5 cm] at (0.25,0.25)  {\scalebox{0.8}{EA-EOM-CCSD}};
        \draw[dashed] (0.25,0.25) -- +(-0.06,0) node [midway] (L) {};
        \node[anchor=west,minimum height = 0.5 cm] at (0.25,0.20)  {\scalebox{0.8}{CCSD}};
        \draw[-] (0.25,0.20) -- +(-0.06,0) node [midway] (L) {};

    \end{scope}
\end{tikzpicture}
    \end{subfigure}
    \begin{subfigure}{0.48\textwidth}
       
        \begin{tikzpicture}
    \node[anchor=south west,inner sep=0] (image) at (0,0) { \includegraphics[width=\linewidth]{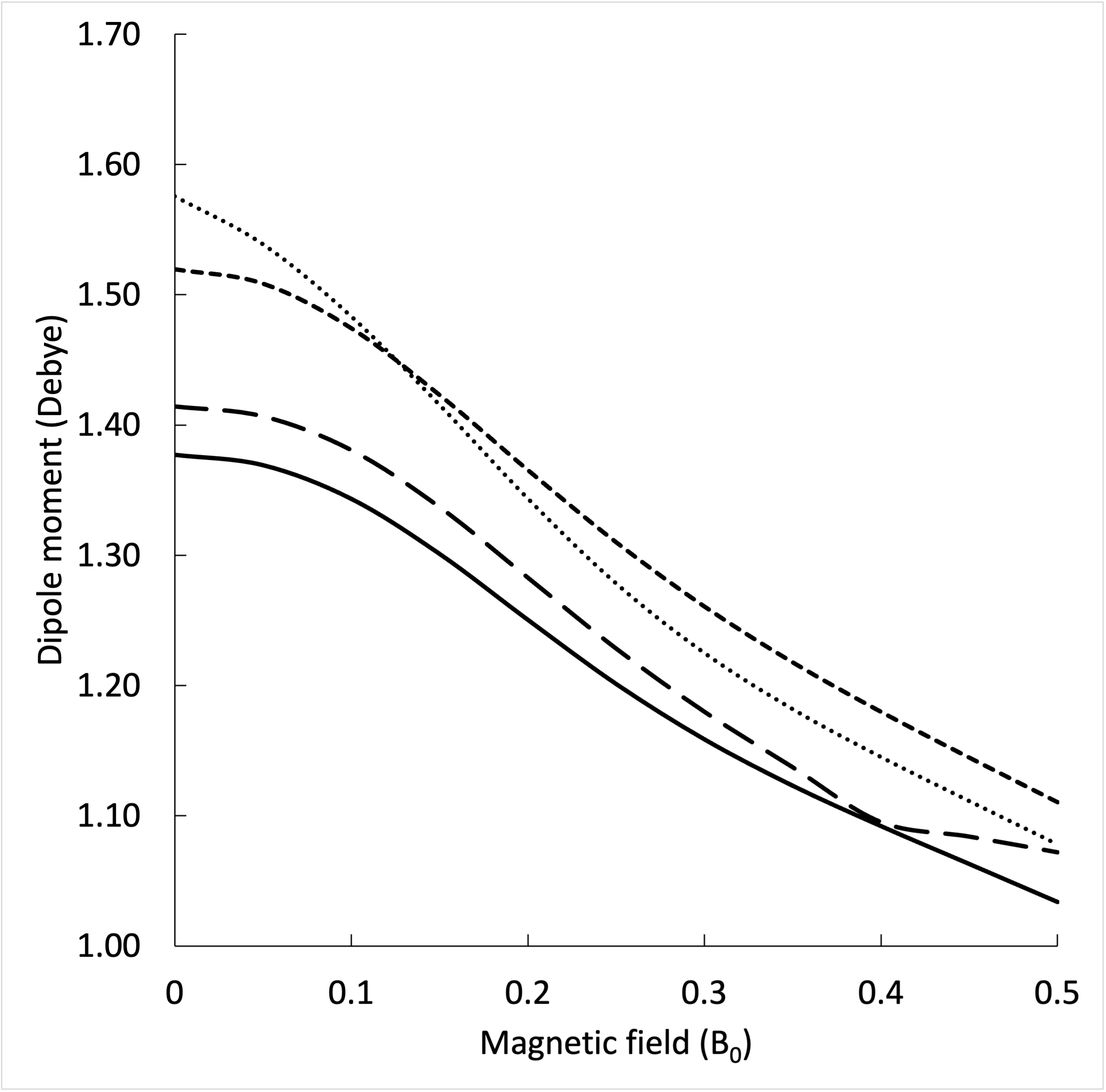}};
    \begin{scope}[x={(image.south east)},y={(image.north west)}]
        \node at (0.25,0.95) (angle) {\scalebox{0.9}{$\phi=30^\circ$}};

    \end{scope}
\end{tikzpicture}
    \end{subfigure}
    \\
    \begin{subfigure}{0.48\textwidth}
        
        \begin{tikzpicture}
    \node[anchor=south west,inner sep=0] (image) at (0,0) {\includegraphics[width=\linewidth]{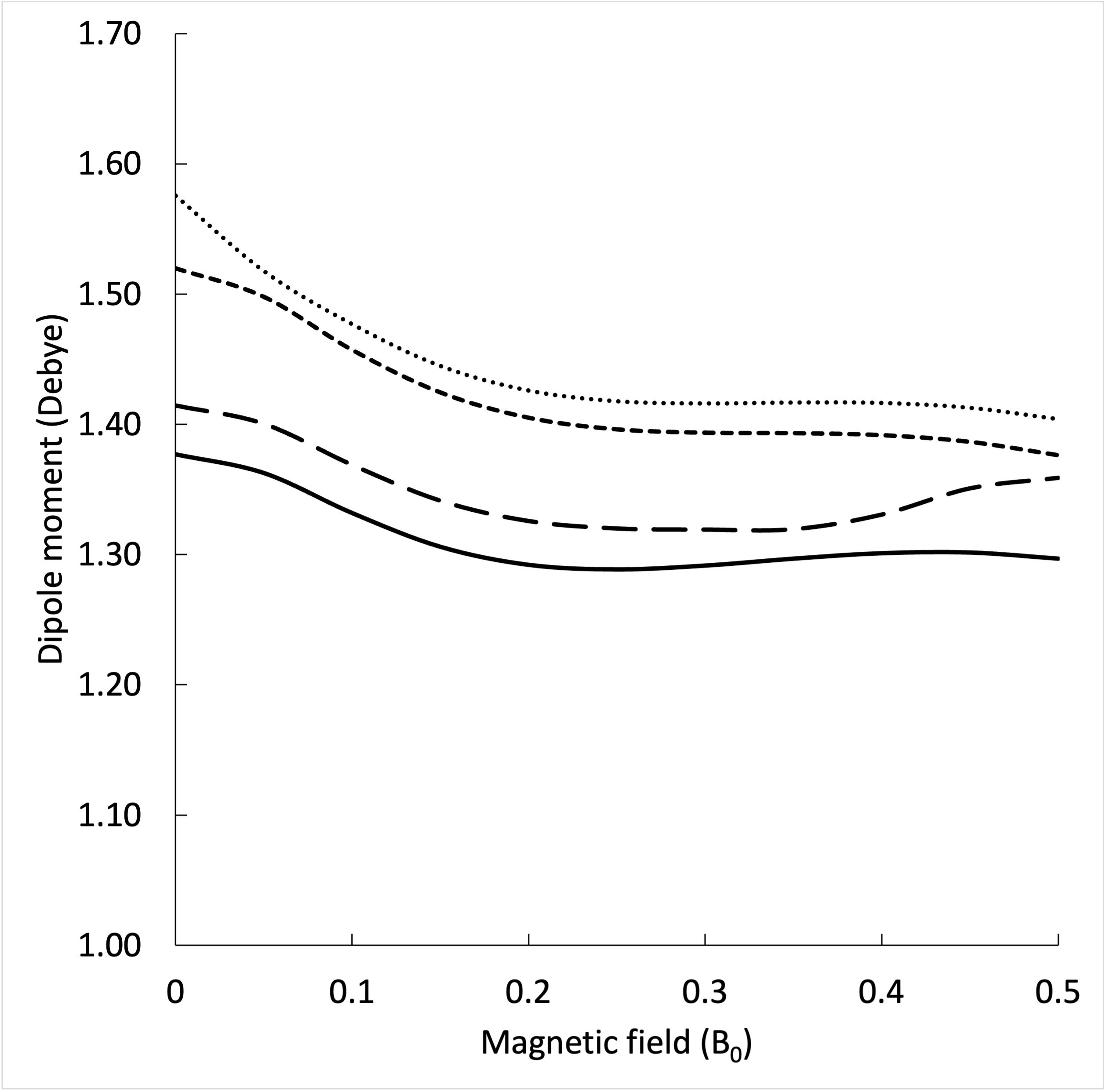}};
    \begin{scope}[x={(image.south east)},y={(image.north west)}]
        
        \node at (0.25,0.95) (angle) {\scalebox{0.9}{$\phi=60^\circ$}};

    \end{scope}
\end{tikzpicture}
    \end{subfigure}
    \begin{subfigure}{0.48\textwidth}
        
        \begin{tikzpicture}
    \node[anchor=south west,inner sep=0] (image) at (0,0) {\includegraphics[width=\linewidth]{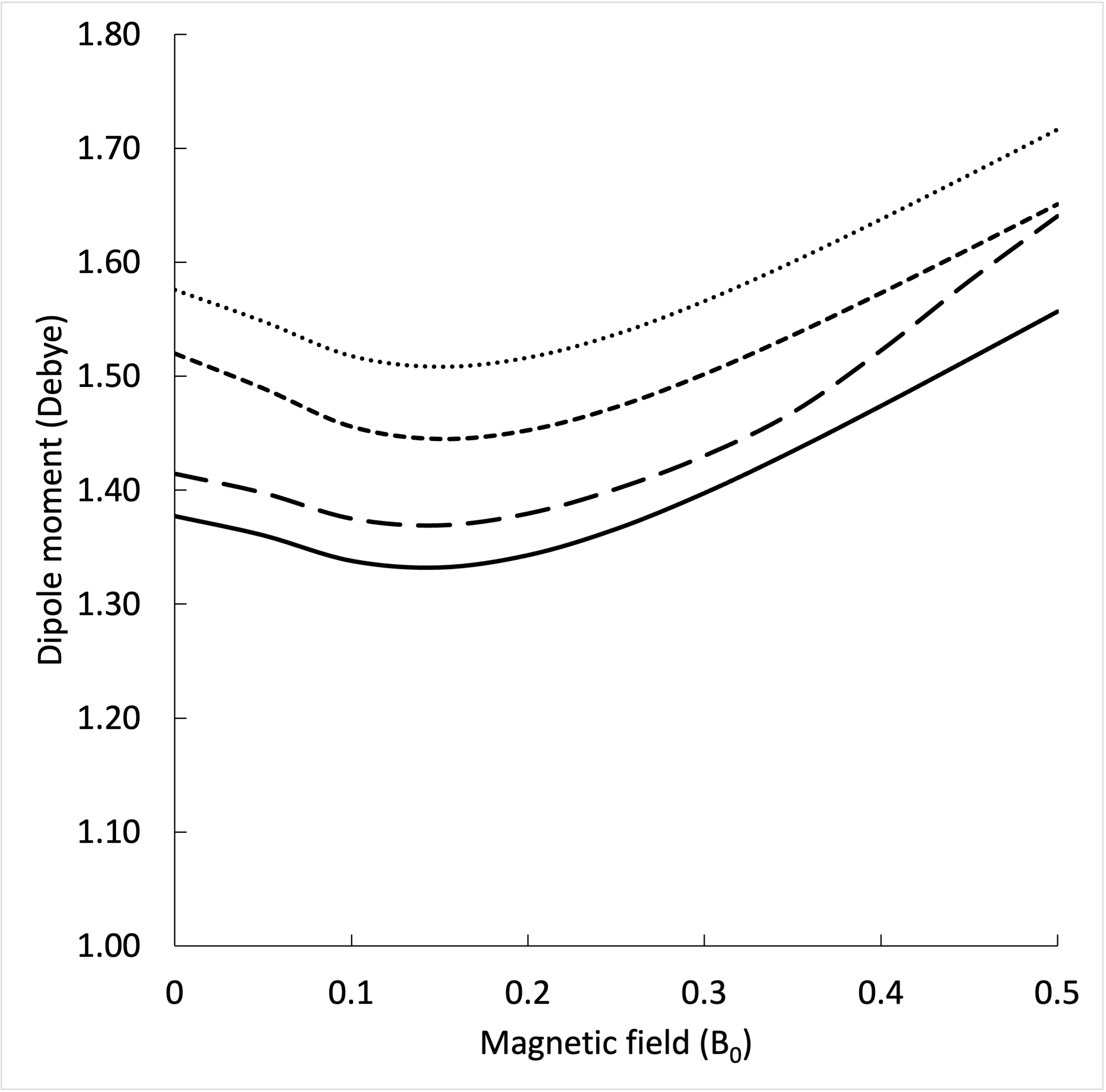}};
    \begin{scope}[x={(image.south east)},y={(image.north west)}]
        \node at (0.25,0.95) (angle) {\scalebox{0.9}{$\phi=90^\circ$}};

    \end{scope}
\end{tikzpicture}
    \end{subfigure}
    \caption{\label{dipmomch}Dipole moment of the CH radical in the lowest doublet state ($^2\Pi$ in the field-free limit) as a function of the magnetic field, calculated at the HF, CCSD, SF-EOM-CCSD, and EA-EOM-CCSD levels of theory for different orientations of the magnetic field with respect to the molecular axis.}
\end{figure}

\subsubsection{Excited states via EA-EOM treatment}
Predictions for the electronic energy for both $^2\Pi$ components using the EA-EOM-CC approach are presented in \figref{fig:CH_EA}. The approach performs very well for most magnetic-field orientations and strengths studied and the EA-EOM-CCSD(T)(a)* results are nearly indistinguishable from their (EE-EOM-)CCSDT counterparts. For the lower-lying state the differences range between $ 0.55  \mEh$ and $ 1 \mEh$, while for the higher-lying state excluding the $30^\circ$ orientation, they lie between $ 0.5 \mEh$ and $ 8  \mEh$.  
Importantly, for an orientation of $30^\circ$, a qualitative difference between the EE-EOM-CCSDT and the EA-EOM-CCSD as well as the EA-EOM-CCSD(T)(a)* results is observed. This is due to the fact that an avoided crossing between the $^2\Pi/2^2A$ and the $^2\Delta/3^2A$ states which occurs at the CCSDT level at $B=0.1\BB$ is missed in the EA-EOM treatments. The avoided crossing is missed since the $^2\Delta/3^2A$ state has a predominant double-excitation character with respect to the reference and is hence shifted away to too high energies when described at the EA-EOM level. As is well known,\cite{Matthews2016} perturbative triples corrections cannot cure this problem. As such, the EA-EOM-CC results show a qualitatively incorrect picture. A similar issue is observed (but is much less acute) in the $60^\circ$ orientation at a magnetic field strength of $B=0.25\tn{-}0.45\BB$. For this region, the maximum energy deviation of the EA-EOM-CCSD(T)(a)* approach relative to CCSDT is $8\mEh$.
While the use of the EA-EOM-CC approach for the higher-lying component of the $^2\Pi$ state is hence not advantageous over the standard EE-EOM-CC approach, 
it yields consistently good results free of spin-contamination and symmetry breaking for the energetically lower component.

\subsubsection{Excited states via SF-EOM treatment}
In \figref{fig:CH_SF}, the energies of the low-lying states calculated at the SF-EOM-CC level of theory are compared against the EE-EOM-CCSDT predictions.  In contrast to the treatment using the standard EE-EOM-CC approach all states of interest have a predominant single excitation character and the assignment of states is much more straightforward. This means that when the character is exchanged between the states, the SF-EOM approach is able to treat them with the same level of accuracy, unlike the EE-EOM approach where the accuracy was significantly compromised when double-excitation character was prominent.\cite{Hampe2020,Kitsaras2024} 

We find that for weaker fields up to $B=0.2\BB$, the deviation of the CCSD(T)(a)* results from those with full inclusion of triples is below $1\mEh$.
Unlike for EE-EOM, using the SF-EOM-CC approach, predictions for the $^2\Sigma^+ $ state (pink) could be obtained in all four orientation studied. 
A significant observation is that contrary to the EA-EOM  results, the two components of the $^2\Pi$ state (red and blue) are well behaved throughout all magnetic-field strengths and orientations studied using the SF-EOM, as well as the EE-EOM approach where one component is the reference state. In addition, the $M_S=-\frac{1}{2}$ component of the $^4\Sigma^-$ state (brown) is well described within the SF-EOM treatment, which is however not the case for EE-EOM.
In the case of the skewed orientations at $30^\circ$ and $60^\circ$, the $^2\Delta/3^2A$ state (yellow) acquires some double-excitation character for $B>0.25\BB$. In these cases,  SF-EOM-CCSD(T)(a)* performs moderately, with an increased deviation from the CCSDT reference of about $1\mEh$. 
For even higher-lying states, issues in the quality of the description are encountered:
For the $^2\Delta/4{^2A}$ (green) and $^2\Sigma^-/5{^2A}$ (purple) states, in the $30^\circ$ magnetic-field orientation, the SF-EOM-CC predictions deviate from  EE-EOM-CCSDT by about $\sim 10\mEh$ for $B>0.3\BB$. This is even more apparent for the $60^\circ$ orientation, where an avoided crossing at $B=0.2\BB$ is completely missed by the SF-EOM-CC predictions. 
Approximate triple corrections at the CCSD(T)(a)* level of theory cannot sufficiently correct the predictions, since they are not designed to account for a predominant double-excitation  character. In the perpendicular orientation, the $^2\Delta/2^2A''$ state (green) is well behaved for all magnetic-field strengths studied, since the mixing with the problematic $^2\Sigma^+/3^2A'$ state (pink) is symmetry forbidden. 
For the $^2\Delta/2^2A'$ (yellow) and $^2\Sigma^-/3^2A''$ (purple) states, the  deviation of the SF-EOM-CCSD(T)(a)* results relative to EE-EOM-CCSDT increases from $\sim0.1\mEh$ to $\sim10\mEh$ for $B>0.25\BB$.

From the discussion above, we note that the energetically lowest $^2\Pi$ component is described accurately with all the approaches, all magnetic-field strengths and all  orientations studied here. The SF-EOM and EA-EOM results with perturbative triple corrections deviate by $\sim 0.1\mEh$ from  the full EE-EOM-CCSDT results. As a comparison, the triples contributions (evaluated using full EE-EOM-CCSDT) vary between $2$ and $3\mEh$.\cite{Kitsaras2024}  EA-EOM-CCSD(T)(a)* recovers $70\tn{-}80 \%$ of the CCSDT triples correction while SF-EOM-CCSD(T)(a)* recovers $60\tn{-}70 \%$. Both EA- and SF-EOM-CC approaches deliver results free of symmetry breaking. The EA-EOM results are furthermore free of spin contamination. SF-EOM-CC is, however, more appropriate for the study of the CH radical, as it can more consistently target the low-lying electronic states of interest beyond the lowest $^2\Pi$ component with the same level of accuracy.

\subsubsection{Dipole moments}
For the lowest doublet state, dipole moments were calculated as EOM-CCSD expectation values and were compared to those obtained at the HF and CCSD levels of theory. The corresponding results are plotted in \figref{dipmomch} as a function of the magnetic-field strength. Despite the close agreement for the energy results using different approaches, the dipole moments are more sensitive to the choice of method. 
Moreover, it is clear that the dipole moment changes quite drastically depending on the the orientation and the magnetic field strength, leading to a qualitatively different behaviour: In the parallel orientation, the dipole moment changes comparatively little, decreasing only slightly from about $1.38$ to $1.28\ \tn{D}$ for the CCSD level of theory when increasing the field strength from $0$ to $0.5 \BB$. In contrast, when the magnetic field axis is tilted to an angle of $30^\circ$, the decrease of the dipole moment is much more pronounced, i.e., going down to about $1.03 \ \tn{D}$.
When changing the angle to $60^\circ$, the steepness of the dipole moment curve as a function of $B$ decreases again, leading to a  
net difference of $0.08 \ \tn{D}$, similar to the parallel case. 
Yet the curvature is quite different: While in the parallel case, the dipole moment decreases throughout, at $60^\circ$ for CCSD, there is a decrease of the dipole moment until about $0.25\BB$, followed by a slight increase up to $0.45\BB$ and a decrease thereafter.
In the perpendicular orientation, the behaviour is similar to the $60^\circ$ case until about $0.15\BB$ after which the dipole moment increases quite significantly, leading to a value of about $1.56 \ \tn{D}$ at $0.5\BB$.
Qualitatively, these trends are reproduced for all methods considered here. Dipole moments computed at the HF, SF- and EA-EOM-CCSD levels are overestimated compared to the CCSD results. While the SF-EOM predictions are closer to CCSD as compared to the other levels of theory, the EA-EOM curves seem to follow the CCSD trends more closely and are parallel to the corresponding CCSD results. The only exceptions are SF-EOM-CCSD results for 30 and 60 degrees where only these results show an increase for strong fields of $0.4 - 0.5\BB$.

For the study of the CH radical, the increased flexibility offered by the different EOM-CC flavours shows merit. The EA-EOM-CC approach manages to describe the energetically lowest $^2\Pi$ state well, without spin-contamination and symmetry breaking. Dipole moments calculated at this level of theory qualitatively agree with CCSD results. However, the EA-EOM results are not consistent in the description of the second $^2\Pi$ component. Furthermore, using the EA-EOM protocol described above, higher lying excited states cannot be well described. The SF-EOM-CC method manages to consistently target all states studied for field strengths $B<0.2\BB$. The inclusion of perturbative triples corrections at the SF-EOM-CCSD(T)(a)* level gives results indistinguishable from EE-EOM-CCSDT predictions as long as the double-excitation character is low. Neither approach is well suited for the description of double-excitation character. The usefulness of SF-EOM should however not be underestimated, especially compared to the use of the standard EE-EOM-CCSD approach which has been proven to be problematic for the study of the excited states of the CH radical.\cite{Kitsaras2024,Kitsarasthes}

\section{Conclusion}
In this paper, the IP, EA, and SF flavors of the EOM-CC approach were implemented as ff-methods. In addition, inclusion of approximate triple excitations at the ff-CCSD(T)(a) and ff-EOM-CCSD(T)(a)* levels of theory were implemented. These approaches were used for studying the IPs and EAs of the lighter elements of the first two rows of the periodic table in the presence of a magnetic field. Exploiting the increased flexibility of the implemented methods, heavier alkali and alkaline-earth metal atoms from the third and fourth row of the periodic table were studied as well. Following their recent discovery in a magnetic WD star, the IPs of Na and Mg, as well as the electronic excitations of Ca were investigated.  Lastly, the EA-EOM-CC and SF-EOM-CC methods were applied to the study of the low-lying electronic states of the CH radical, a molecule of interest for  magnetic WD stars. 

The development of the IPs and EAs to an increasing magnetic-field strength is dominated by the Landau energy of the free electron that is ejected or captured, respectively. The paramagnetic interaction does not contribute to the development. The diamagnetic interaction of the free electron scales as $B$ while the electronic diamagnetic interaction scales as $B^2$. For the magnetic fields studied, the diamagnetic Landau energy dominates and as such,  IPs are destabilized while EAs are stabilized when increasing the field strength. 
The deciding factors, however, are the ground state of the system and the character of the captured/ejected electron. Changes of these lead to discontinuities of the IP/EA as a function of the magnetic field strength or their derivatives, i.e., slopes. 

The electronic structure of Na, Mg, and Ca was investigated. Firstly, calculation on the IPs of Na and Mg reveals that the diamagnetic contribution is more prominent when compared to the IPs of the lighter elements, due to the larger size of the former. Moreover, studying the electronic triplet states of Ca in the presence of a magnetic field showed qualitative differences as compared to Mg. The low-lying $^3D_g$ state that arises from the $3d$ orbitals leads to a different ground state of Ca compared to the lighter Mg in stronger magnetic field strengths. Nonetheless, the electronic excitation studied, i.e. the $4p\rightarrow5s$ transition of Ca is very similar to the respective excitation of Mg. The fact that Ca belongs to the fourth row of the periodic table means that full inclusion of all-electron triple corrections at the EOM-CCSDT level is not feasible. As such, the newly implemented EOM-CCSD(T)(a)* approach is significant for flexibly and accurately studying the system using the SF flavor. 

Studying the CH radical with the increased flexibility of the non-standard EOM-CC variants shows remarkable advantages. Firstly, using the EA-EOM-CC flavor allows the targeting of the ground-state of the system in a way free of spin-contamination  and symmetry-breaking. Secondly, the low-lying states of the system can be accessed easily with a predominant single-excitation character via the SF-EOM-CC approach. While consistency is not fully achieved for all the magnetic-field strengths and orientations studied, the SF-EOM-CCSD variants is a significant improvement compared to the EE-EOM-CCSD approach.  Moreover, the SF-EOM-CC approach gives results free of symmetry breaking, which facilitates the characterization of the states. 

The results show, that the IP, EA, and SF-EOM-CC approaches are advantageous for a careful study of complex electronic structures in the presence of a magnetic field. Approximate triple corrections at the EOM-CCSD(T)(a)* level have the potential to give highly accurate results for the assignment of spectra from magnetic WDs. In particular, it is the flexibility offered by having access to various ff-EOM-CC flavors that enables the accurate treatment of excited states. At the same time, the fact that the character of the excitation can change over the range of different field strengths as well as orientations remains a challenge in the generation of reliable $B-\lambda$ curves.

\section{Acknowledgements}
The authors thank  Jürgen Gauss for fruitful discussions.  This work has been supported by the Deutsche Forschungsgemeinschaft under the grant STO 1239/1-1.
The authors are grateful to the Centre for Advanced Study at the Norwegian Academy of Science and Letters, Oslo, Norway, where early parts of this work were carried out under the project “Molecules in Extreme Environments”.

\printbibliography

\end{document}